\documentclass[journal,draftcls,onecolumn,12pt,twoside]{IEEEtranTCOM}
\usepackage{amsmath}
\usepackage{epsfig}
\usepackage{graphicx}
\usepackage[font={small},labelfont=small, bf]{caption}
\usepackage[subfigure]{tocloft}
\graphicspath{{image/}}
\usepackage[table,xcdraw]{xcolor}
\usepackage[utf8x]{inputenc}
\usepackage{caption}
\usepackage{algorithmic}
\usepackage{subfloat}
\usepackage{float}
\usepackage{amsthm,amssymb,amsmath,bm}
\usepackage{subfigure}
\usepackage{amsfonts}
\usepackage{epsfig}
\usepackage{amssymb}
\usepackage{amsmath}
\usepackage[table,xcdraw]{xcolor}
\usepackage{cite}
\usepackage{multicol}
\usepackage[utf8x]{inputenc}
\usepackage{color,soul}
\usepackage{subfigure}
\usepackage{multirow}
\usepackage{rotating}
\usepackage{graphicx}
\usepackage{tabularx}
\usepackage{array}
\usepackage{color,soul}
\usepackage{yfonts,color}
\usepackage{siunitx}
\usepackage{lettrine}
\usepackage{bm}
\usepackage{graphicx,dblfloatfix}
\usepackage{blindtext}

\usepackage{subfigure}
\usepackage{moreverb}
\usepackage{epsfig}
\usepackage{amsmath,amssymb,amsthm,mathrsfs,amsfonts,dsfont}
\usepackage{adjustbox,lipsum}
\usepackage{amsfonts}
\usepackage{epsfig}
\usepackage{amssymb}
\usepackage{amsmath}
\usepackage{amsthm}
\usepackage{subfigure}
\usepackage{multirow}
\usepackage{rotating}
\usepackage{graphicx}
\usepackage{tabularx}
\usepackage{array}
\usepackage{anyfontsize}
\usepackage{color,soul}
\usepackage{graphicx,dblfloatfix}
\usepackage{epstopdf}
\usepackage{blindtext}
\usepackage{amsmath}
\usepackage{amsthm,amssymb,amsmath,bm}
\usepackage{subfigure}
\usepackage{amsfonts}
\usepackage{epsfig}
\usepackage{amssymb}
\usepackage{amsfonts}
\usepackage{url}
\usepackage{amsmath}
\usepackage{cite}
\usepackage{graphicx}
\usepackage{fancyhdr}
\usepackage{subfigure}
\usepackage[subfigure]{tocloft}
\usepackage[font={small}]{caption}
\usepackage{subfigure}
\usepackage{tabularx}
\usepackage{tcolorbox}
\usepackage{cite}
\usepackage[linesnumbered,ruled,vlined]{algorithm2e}
\SetKwInput{KwInput}{Input}
\SetKwInput{KwOutput}{Output}
\usepackage{amsthm,amssymb,amsmath,bm}
\usepackage{graphicx}
\usepackage{fancyhdr}
\usepackage{subfigure}
\usepackage[subfigure]{tocloft}
\usepackage[font={small}]{caption}
\usepackage{subfigure}
\usepackage{tabularx}

\usepackage{cite}
\allowdisplaybreaks
\usepackage[colorlinks,bookmarksopen,bookmarksnumbered,citecolor=red,urlcolor=red]{hyperref}

\begin{document}
	\author{\IEEEauthorblockN{Marjan Tajik, Mohammadreza Maleki, Nader Mokari, \IEEEmembership{Senior Member, IEEE}, Mohammad Reza Javan,  \IEEEmembership{Senior Member, IEEE}, Hamid Saeedi , \IEEEmembership {Member, IEEE}, Bile Peng, \IEEEmembership {Member, IEEE}, and Eduard~A.~Jorswieck, \IEEEmembership{Fellow, IEEE}
			%}}
			\thanks{  
				M.Tajik, M. R. Maleki, 
				N.~Mokari, and H.~Saeedi are with the Department of Electrical and Computer Engineering,~Tarbiat Modares University,~Tehran,~Iran, (e-mail: {t.marjan, M.mohammadreza, nader.mokari, hsaeedi}@modares.ac.ir ). M.~R.~Javan is with the Department of Electrical Engineering, Shahrood University of Technology, Iran, (e-mail: javan@shahroodut.ac.ir). Bile Peng and Eduard A. Jorswieck are with TU Braunschweig, Department of Information Theory and Communication Systems, Braunschweig, Germany (e-mail: {peng, jorswieck, jorswieck}@ifn.ing.tu-bs.de).
	}}}
	\title{ \huge Two-Hop Age of Information Scheduling for Multi-UAV Assisted Mobile Edge Computing:\\  FRL vs MADDPG }
	%%\author{\IEEEauthorblockN{n}\\
	%}
	\maketitle
	\markboth{IEEE Transactions on Communications}%
	{Submitted paper}
	\begin{abstract}
		In this work, we adopt the emerging technology of mobile edge computing (MEC) in the Unmanned aerial vehicles (UAVs) for communication-computing systems, to optimize the age of information (AoI) in the network. 
		We assume that tasks are processed jointly on UAVs and BS to enhance edge performance with limited connectivity and computing. Using UAVs and BS jointly with MEC can reduce AoI on the network.
		To maintain the freshness of the tasks, we formulate the AoI minimization in two-hop communication framework, the first hop at the UAVs and the second hop at the BS.
		To approach the challenge, we optimize the problem using a deep reinforcement learning (DRL) framework, called federated reinforcement learning (FRL). In our network we have two types of agents with different states and actions but with the same policy. Our FRL enables us to handle the two-step AoI minimization and UAV trajectory problems. In addition, we compare our proposed algorithm, which has a centralized processing unit to update the weights, with fully decentralized multi-agent deep deterministic policy gradient (MADDPG), which enhances the agent's performance. As a result, the suggested algorithm outperforms the MADDPG by about 38\%. 
		%Our goal is the timeliness of MEC systems in which information freshness is essential.
		%such as the industrial internet of things (IIoT), augmented reality (AR), and virtual reality (VR). 
		%The simulation result shows that the average AoI using partial computing is smaller, while the number of devices is high in the network.\\
	\end{abstract}
	\section{Introduction}
	\subsection{Motivation}
	%Fifth-generation (5G) and sixth-generation (6G)
	  New generation wireless networks need to support a broader range of services such as low latency and reliable communication, massive communications, and enhanced mobile broadband. Consequently, traditional networks are forced to fulfill the requirements of all these various services \cite{9285214}\cite{abedin2018resource}. The unmanned aerial vehicles (UAVs) can provide feasible solutions for new generation networks and they are used as an aerial base station (BS) for gathering and transmitting data. Basically, the UAVs have emerged as an advanced technology with low cost, flexible mobility, and direct line-of-sight (LoS) communication with ground infrastructures \cite{li2018uav}. The UAVs are able to perform computationally-intensive and latency-critical tasks in the future; however, in practice, they are limited by the amount of energy, weight, and space \cite{mozaffari2019tutorial}\cite{li2018uav}. Therefore, designing UAV networks to meet their different service requirements is challenging. We exploit mobile edge computing (MEC) techniques to address these challenges. The concept of MEC has evolved in recent years and has provided computing capability at the edge of wireless networks. MEC in high-mobility UAVs creates numerous opportunities and challenges in determining the optimal method for offloading computation. MEC provides the ability to perform computations close to the user, which reduces the energy consumption and the amount of time required to perform computationally intensive tasks. 
	  As a result, the MEC implementation is designed either to be at the BS or other edge sites in order to provide customers with a rapid and flexible deployment of new applications and services.
	The amount of data to be offloaded to the MEC must be determined by factors influencing the network. The topic of edge processing in the UAVs has recently been raised, and it is expected to be more flexible and faster than the conventional MEC. UAV-aided MEC deployments in large-scale internet of thing (IoT) scenarios can increase users quality of service (QoS) by offloading some computing tasks to the UAV's MEC \cite{yang2020multi}.
	In addition, a wide range of applications, for instance, mobile online games, face recognition, and augmented reality (AR), require a high performance and low latency network, which MEC can provide \cite{mao2017survey}.\\
	Several real-time applications have strict requirements in terms of freshness of status, and the information must be successfully delivered on time \cite{chen2020minimizing}. In these applications, timely information awareness is an important requirement for real-time monitoring and control. The users rely on the computation results on time to make the right decisions.
	To maintain the freshness of data at the destination, a new metric called the age of information (AoI) was introduced. AoI is defined as the time elapsed since the source’s latest update of received information status at the destination \cite{yates2021age}. Data staleness can be measured by applying the AoI from the destination viewpoint \cite{chen2020multiuser}. Therefore, ensuring the QoS in AoI is necessary for each time-critical task. AoI minimization is fundamentally different from delay minimization, or throughput maximization \cite{abdel2018ultra}. In some services such as telemedicine and safety-related event broadcasting, ensuring a tolerable AoI is critical. In this case, the out-of-data measurement may lead to incorrect decisions since the data is incompatible with the current state of the target. 
	{\color{black} Also, in some real-time processes, where the effect of task processing on AoI is unavoidable, MEC can be used for decreasing the AoI of computationally-intensive tasks.} This paper also assumes BS's are  equipped with MEC to address the UAVs’ limited capacity and power for processing. By processing at different edges of the network, we can improve AoI and reduce processing time.
	\subsection{Related works}
	In this subsection, we discuss the related works in various categories, such as MEC in UAV, AoI with MEC, and AoI without MEC.
	\subsubsection{MEC and UAV}
	In \cite{ei2021multi}, the authors jointly optimize device association, content assignment, and resource allocation to minimize the energy consumption of mobile devices in multi-UAV assisted MEC computing systems. An iterative block coordinate descent (BCD) algorithm is used in this study to solve the formulated problem after decomposing it into three subproblems. A UAV assisted MEC system is presented in \cite{alsenwi2020uav}, with the aim to minimize the energy consumption of IoT devices, including the energy used in local computation and uplink transmission, as well as the energy consumed by UAVs. Therefore, they develop a BCD algorithm for offloading tasks, allocating bandwidth, assigning local computation resources, and UAV computation resources. 
	The authors in \cite{yang2019energy} minimize the total power by jointly optimizing power control, user association, computing capacity, and flight trajectory in the UAV network with MEC.
	Due to the processing limitations in IoT devices, multiple UAVs act as a MEC server in the network for some processing jobs \cite{yang2020multi}. A multi-UAV deployment mechanism based on differential evolution (DE) is proposed to balance the load on the UAVs. Furthermore, a deep reinforcement learning (DRL) is developed to schedule a particular UAV. According to \cite{peng2020ddpg}, the authors study joint vehicle association and resource management in a vehicular network using macro eNodeB (MeNBs) and the UAVs, which are equipped with MEC, and they also use a deep deterministic policy gradient (DDPG) based algorithm. Furthermore, a UAV-based MEC framework is described in \cite{wang2020multi}, in which several UAVs with different trajectories fly over the target area, and support the UEs on the ground. The authors propose a multi-agent DRL-based trajectory control algorithm that is optimizing geographical fairness among users, energy consumption, and UE-load for each UAV.	
	\begin{table*}
		\centering
		\caption{Previous works}
		\label{last}
		\scalebox{0.7}{
			\begin{tabular}{|cccccl|}
				\hline
				Refrence & Multi-UAV & MEC       & Trajectory Design & Objective Function      &  Solution                                        \\ \hline
				\cite{9285214}        & $ \surd $ & $\times$  & $\surd$       &Maximize EE under AoI constraint & DQN                          \\ 
				%			\cite{alsenwi2020uav}        & $ \surd $ & $ \surd $ & $\times$      & Energy consumption                                  \\ 
				%			\cite{ei2021multi}         & $ \surd $ & $ \surd $ & $\times$      & Local computing energy consumption                  \\ 
				\cite{hu2020cooperative}       & $ \surd $ & $\times$  & $ \surd $     & Minimize the accumulated AoI   & CA2C, DDPG, and DQN                   \\ 
				\cite{li2019minimizing}       & $\times$  & $\times$  & $ \surd $     & Minimize the total number of expired packets and minimize AoI     & RL                   \\ 
				%\cite{peng2020ddpg}       & $ \surd $ & $ \surd $ & $\times$      & Number of task successfully completed by MEC server \\ 
				\cite{tong2019uav}        & $\times$  & $\times$  & $ \surd $     & Minimize AoI    &     DP             \\ 
				%\cite{wang2020multi}       & $ \surd $ & $ \surd $ & $ \surd $     & Fairness of each UE-load and overal Energy          \\ 
				\cite{8570843}    &$\times$  & $\times$  &  $ \surd $ & Minimize PAoI & Iterative algorithm  \\
				\cite{yi2020deep}        & $\times$  & $\times$  & $ \surd $     & Minimize AoI & DRL \\
				\cite{liu2018age}      &$\times$  & $\times$  &  $ \surd $ & Minimize AoI & DP, GA  \\
				\cite{hu2020aoi}       &$\times$  & $\times$  &  $ \surd $ & Minimize the AAoI &KKT, DP, Ant Colony \\
				\cite{samir2020age}     & $ \surd $  & $\times$  &  $ \surd $ & Minimize the Expected Weighted Sum AoI &DRL, DDPG \\
				\cite{wu2021uav}       & $ \surd $ & $\times$  & $ \surd $     & Minimize AoI & MADRL, DDPG                                               \\ 
				%\cite{yang2020multi}       & $ \surd $ & $ \surd $ & $ \surd $     & Balancung the load for UAVs                         \\ 
				\cite{zhang2020age}       & $\times$  & $\times$  & $ \surd $     & Minimize AoI  & RL, DP                           \\ 
				\cite{zhou2019deep}       & $\times$  & $\times$  & $ \surd $     & Minimize AoI     &        DRL                                    \\ 
				\cite{zhu2021federated} &   $ \surd $ & $ \surd $ &$\times$     & Minimize AoI    & Hetrogenous multi agent actor-critic (FRL)                    \\ 
				\cite{han2021age} &   $\times$  & $ \surd $ &  $\times$  & Minimize AoI    & Markov chain                     \\ 
				\cite{choudhury2021aoi} & $ \surd $ &  $\times$  &   $\times$  & Minimize AoI    &DQN                     \\
				Our work  & $ \surd $ & $ \surd $ & $ \surd $     & Minimize AoI         & FRL, MADDPG                                        \\ \hline
			\end{tabular}
		}
	\end{table*}	
	\subsubsection{AoI without MEC}
	In previous works, the authors optimize AoI in UAV networks without MEC, and most of the papers focus on trajectory design. In \cite{li2019minimizing}, the authors consider a UAV trajectory planning pattern to minimize expired data packets from all sensors and then relax the unclear original problem into a Min-Max-AoI optimal path design, which is based on reinforcement learning (RL). In \cite{8570843}, the authors consider the UAV as a mobile relay and formulate the average peak AoI (PAoI) minimization problem to optimize the UAV's flight trajectory, energy allocations, and periods for transmitting update packets in the source and the UAV. To solve this non-convex problem, they suggest an effective iterative algorithm and prove the convergence analytically. In \cite{liu2018age}, the authors investigate age-optimal trajectory planning in wireless sensor networks using UAVs to collect data from the ground sensor.
	% AoI gathered from each SN is affected by data uploading time and the length of time since the UAV left SN and which gathering data from other SNs.
	{\color{black} There are two trajectory plans considered, Max-AoI and Ave-AoI}. A genetic algorithm (GA) and dynamic programming (DP) approach are used to determine the age-optimal trajectory.
	Moreover, in \cite{yi2020deep} and\cite{abd2019deep}, the UAV's trajectory and the transmission schedule are optimized to achieve the minimum AoI. In \cite{yi2020deep}, the authors have maintained the freshness of information in IoT networks using UAVs. The UAV collects status update packets during its flight and maintains non-negative residual energy while flying towards the sensors. They propose DRL algorithm to overcome this problem.
	In \cite{samir2020age}, the authors discuss UAV-assisted single-hop vehicular networks. They use DDPG for finding UAV trajectory and scheduling under minimal throughput requirements, while the deployed UAVs adjust their speeds during data gathering in order to minimize the AoI.
	In \cite{zhou2019deep}, an online UAV trajectory planning is derived in IoT networks with unknown traffic patterns, which minimizes AoI in a network. The trajectory design problem is formulated for the internet of UAVs as Markov decision process (MDP) to minimize AoI \cite{hu2020cooperative}. The authors develop a method to recognize compound-action actor-critics by a distributed sense-and-send protocol.
	In \cite{kadota2018optimizing}, the authors consider average AoI by examining throughput of networks and the minimum required throughput. They apply different policies, such as random and Max-Weight. In \cite{9285214}, the authors create a navigation method for multiple UAVs where BSs are utilized to enhance the data freshness and connection to IoT devices. This work maximizes the energy efficiency of UAV networks under the AoI constraint and AoI thresholds for each user,  by using the DRL algorithm. 
	In \cite{hu2020aoi}, the authors consider a UAV-assisted wireless powered IoT network, where a UAV is launched from a data center, flies to each of the ground SNs, and gathers data, then returns to the data center. 
	The optimization problem is formulated to minimize the average AoI of the data collected from all SNs, which depends on the UAV's trajectory, the time required for energy harvesting (EH), and data collection for each SN. The method used is DP and Ant Colony (AC) heuristic algorithms. A UAV to device (U2D) communication, underpinning the cellular internet of UAVs is discussed in \cite{wu2021uav}, in which the authors design a cooperative sensing and transmission protocol. The goal is to minimize AoI through trajectory design, using multi-agent DDPG algorithm. Similary, \cite{zhang2020age} also minimizes AoI by optimizing cooperative sensing and transmission time, UAV trajectory, and task scheduling. In \cite{tong2019uav}, the authors consider joint sensor association and UAV's flight trajectory, which minimizes AoI. This paper uses DP and Affinity Propagation (AP) based algorithms. 
	\subsubsection{AoI with MEC}
	The authors in \cite{zhu2021federated} focus on the timeliness of the MEC systems, where data freshness and computation tasks are important factors. This model is based on age-sensitive MECs, and minimizes AoI problems. A new multimodal DRL framework is suggested, called heterogeneous multi-agent actor-critic (H-MAAC), which is used to facilitate collaboration among edge devices and the central controller.
	%By allowing connected vehicles to make use of fog computing, AoI can be reduced; the authors in \cite{chen2020minimizing} evaluated the impact of three main factors: update frequency, choice of cloud and fog servers, and processing delay at fog/cloud servers, on the empirical PDFs of AoI in a vehicular system.
	% In \cite{kuang2020analysis}, the focus was on the average AoI for computation-intensive messages in a MEC system and the AoI performance with three different computing strategies, including local computing, remote computing, and partial computing.\\
	%In \cite{chen2020age}, are studied the issue of resource awareness with AoI in a UAV-MEC network deployed by an infrastructure provider (InP). In this work, an online DRL approach, which would allow each mobile user to only use its local assumptions, was considered to approximate the Nash equilibrium solutions.
	Researchers in \cite{han2021age} develop an IoT network utilizing UAVs with MEC where the efficiency of data gathering was determined by measuring packet loss rate and total data quantity using a Markov chain. Also, the computation frequency of UAVs is designed according to the preferred cost coefficients for energy and time consumption in order to meet the diverse requirements of services. AoI is also used to measure the freshness of data packets. Table \ref{last} summarizes the main differences between our paper and related works in the last two categories.
	In addition to considering scheduling policies for AoI optimization in two-hop, which is mentioned in \cite{choudhury2021aoi}, our paper focuses on edge processing capability and minimizing AoI in both hops. Unlike most articles, we optimize AoI before reaching the destination and consider the effect of different delays too. Moreover, due to the limited capacity of the UAV, we do not constrain the processing tasks to be performed only at the UAV.
	%This paper\cite{choudhury2021aoi} considered scheduling policies for AoI optimization in two-hop UAV-relayed IoT systems, but in addition mentioned, our purposed paper focus on minimizing AoI in two hop. The difference is that we considered edge processing capability in both hops, and we considered edge processing capability in both hops an
	% We summarize the difference between our work and the previous works, which focused on AoI in Table. I.
	%%In [], the autrhors focus on the timeliness of the MEC systems where the freshness of the data and computation tasks is significant. they formulate a kind of age-sensitive MEC models and describe the AoI minimization problems of interests. Then, a novel mixed-policy based multimodal deep reinforcement learning (RL) framework, called heterogeneous multi-agent actor-critic (H-MAAC), is proposed as a paradigm for joint collaboration in the investigated MEC systems, where edge devices and center controller learn the interactive strategies through their own observations.
	\subsection{Contribution}
	%In this paper, we discuss resource management in UAVs and BS equipped with edge processing. Real-time resource allocation can be achieved by using multi-agent DRL algorithms. The main contributions of this paper are summarized as follows.\\
	{\color{black} We propose an AI approach to solve the trajectory, task offloading, and resource allocation problem in a UAV-enabled MEC system}. For the first time, we formulate AoI optimization for this problem by considering partial offloading in the UAVs and the BS. Our main contributions are as follows:\\
	$ \bullet$ For time-sensitive services and large-scale networks, we  propose a new architecture. In this regard, we suggest a two-hop MEC-enabled UAV network at which we should guarantee the freshness of Information. To this end, we devise a new formula for AoI in which we focus on the transmission and processing times at the links between the BS and the UAVs and also between the UAVs and the devices.
	In order to minimize AoI, we consider new constraints satisfying each task transmitted and processed at a one-time slot.\\
	%Considering the growing number of time-sensitive and processing-intensive services, we discuss how to minimize AoI in two-hop by processing at the edge of the network. We consider collaboration between two types of edge nodes (UAVs and BS) for partial processing. \\  
	%$ \bullet$ Given that the formulated problem is nonconvex and the complexity of the issue is increasing. Since there are two types of agents in the model we propose the RL method and design a federated learning in this work. 
	$ \bullet$ We exploit multi-agent deep deterministic policy gradient (MADDPG) and federated reinforcement learning (FRL) methods to solve optimization problems. Our system evaluates a novel approach using two types of agents with different states and actions but the same goal. Our FRL enables us to handle the two-step age minimization and UAV trajectory problems. In addition, we present our proposed algorithm with a fully decentralized MADDPG agent that is deprived of the weights update center, which enhances the agent's performance. Also, we investigate the problem in terms of convergence and complexity.\\
	$\bullet$ Due to information sharing in the FRL method, simulation results show that this method performs on average 38\% better than the MADDPG. Also, when network traffic increases, the use of MEC in UAVs and BS at the same time reduces network processing load and increases AoI performance by approximately 28\%. 
	\subsection{ Paper Organization}
	The outline is summarized as follows. In Section II, the system model and problem formulation are discussed. In Section III, we propose the solution to the optimization problem. In Section IV, we transform the formulated problem using RL and resolve it with The FRL algorithm.  simulation results are presented in Section V. In Section VI, we conclude the paper.
	\section{system model}
	\subsection{Overview}
\begin{figure}[ht!]
		\centering
		\includegraphics[width=0.6\linewidth]{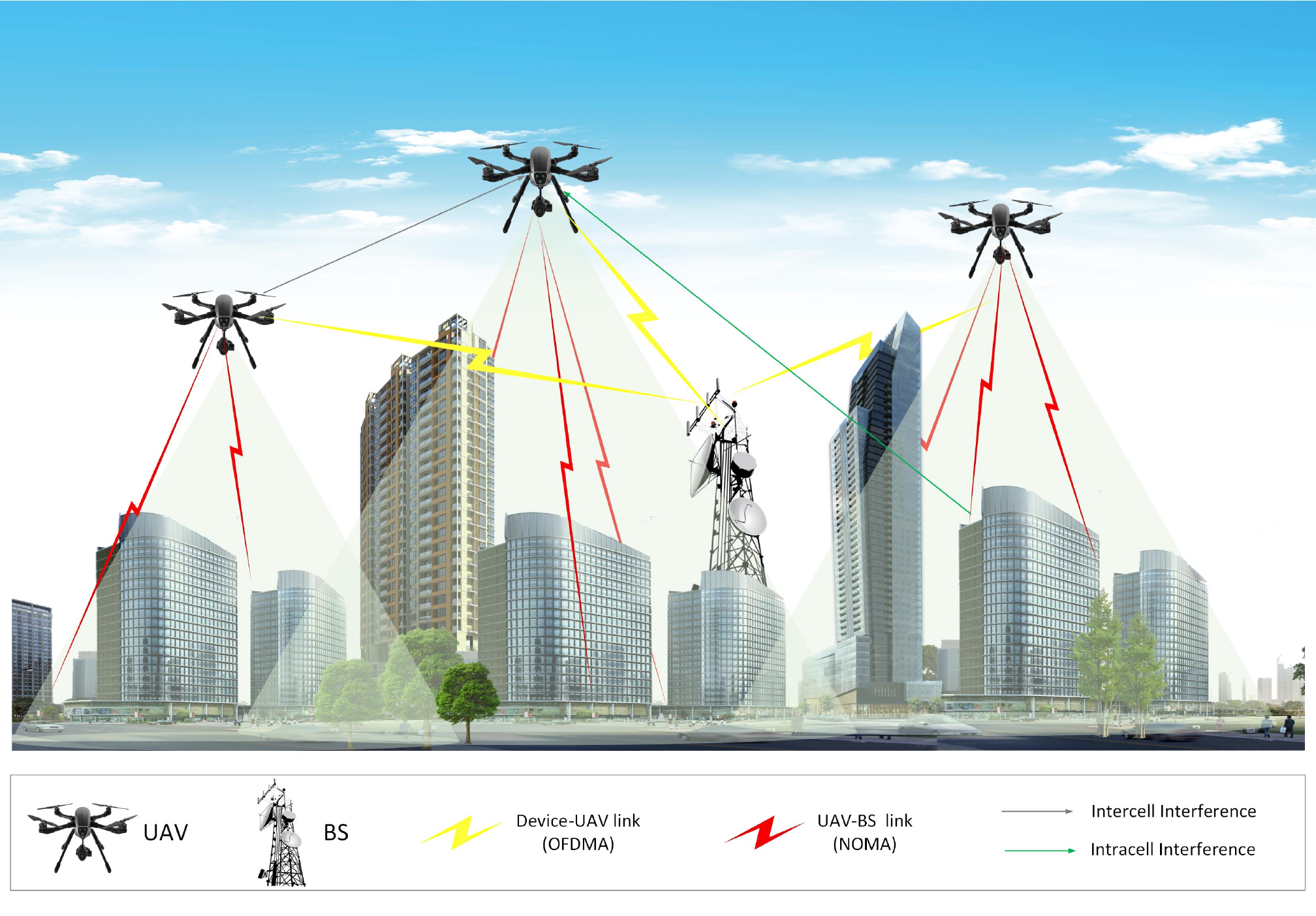}
		\caption{system model.}
		\label{fig 1}
	\end{figure}

	We present a UAV-enabled MEC system scenario as shown in Fig. \ref{fig 1}.
	Assume there is a set of $\mathcal{K} = \{1, . . . , K\}$ devices, which transmit the tasks to a set of $\mathcal{M}= \{1, . . . , M\}$ UAVs for some processing task. Next, the UAV's transmit results to the BS for the remaining processing. Each device is placed uniformly at random. Assume that all the UAVs are at the same altitude $ h_m=h$ at their position $X_{m,t} = [x_{m,t}, y_{m,t}, h_m]$, and the position of each device is {\color{black} $\hat{X}_k = [\hat{x}_k, \hat{y}_k, 0]$}. 
	For better understanding, we summarize symbols and variables in Table \ref{not}.	
	We consider $T$ time slots in the network with one millisecond duration, represented by $t \in \{1, . . . , T\} $. {\color{black} When a previous task is sent to and processed by the BS, a new task is generated}. Each task is denoted by the function $S_{k}=(D_{k},F_{k}), \forall k \in \mathcal{K} $, where $D_{k}$ represents the input data size (in bits) and $F_{k}$ shows the required CPU cycles to execute one-bit data of this task \cite{wang2021deep}. 
	\begin{table}[]
		\centering
		\caption{Main notations}
		\label{not}
		\scalebox{0.7}{
			\begin{tabular}{|ll|ll|}
				\hline
				Notation & Definition  & Notation &     Definition                               \\ \hline
				$\mathcal{K/M/N}$    & Number of devices/UAVs/time & $T_{m,t}^{\text{MEC}}$     &Execution time of UAV $m$ at $t$ [ms] \\
				$X_k/X_m$            & Device/UAV location         &                            	$T_{m,b,t}^{\text{MEC}}$         & Execution time of UAV $m$ in BS at $t$ [ms]               \\
				$R_c$                & UAV radius [m]   & $T_{m,b,t}^{\text{tr}}$        & Total transmission time from UAV $m$ to BS at $t$ [ms] \\ 
				$D_{k}$              & Data valum of device $k$ [bits]    & 		$T_{k,m,t}^{\text{tr}}$       & Data offloading time for device $k$ to UAV $m$ at $t$ [ms]     \\
				$F_k$                & CPU cycle of device $k$ [GHz] & $ f_{m,t}^{\text{MEC}}$    & Computational capability of UAV $m$ at $t$  [GHz]  \\  
				$g_{m,k,t}$         & Channel gain between device $k$ and UAV $m$ at $t$& 	$ h_{m,b,t}^{l}$      & Channel coefficient between UAV $m$ and BS at $t$  \\ 
				$F/L$                  & Number of subcarriers between devices and UAVs/UAVs and BS & $p_{m,,t}^{l}$       & Transmit power of UAV $m$ at $t$ [dBm] [dBm]  \\ 
				$R_{m,k,t}^{f}$     & Data rate between device $k$ and UAV $m$ in subcarrier $f$ at $t$ [bits/ms]& $\zeta_{m,BS}^{l} \in\left\lbrace   0,1\right\rbrace  $   & Allocation of subcarrier $l$ to UAV $m$  \\ 
				$R_{k,t}$           & Total data rate of device $k$ at $t$ [bits/ms]  & $(\sigma_{m}^{l})^2$ & Noise power of UAV $m$ on subcarrier $l$   \\ 
				$ \rho_{k,t}$          & Transmission power of device $k$ at $t$ [dBm]    & $\gamma_{m,b,t}^{l}$  & SINR of  UAV $m$ to BS on subcarrier $l$    \\ 
				$\sigma^2$           & White Gaussian noise variance  & $\psi_{k,f,t} $     & Binary variable for subcarrier $f$ assign to device $k$ at $t$       \\ 		
			    $\Delta_{k,t}^{s}$ & AoI of the processed status device $k$ at BS [ms] &$\tilde{R}_{m,t}$   & Sum transmit rate of  UAV $m$ at time $t$  [bit/ms]   \\ 			
				$\Delta_{k,t}^{p}$           & AoI of the processed status device $k$ at UAV [ms] &$ \lambda_{k} \in \left( 0,1\right) $       & Ratio for the offloaded tasks $k$ to UAV [bit/ms]                    \\ \hline               				
			\end{tabular}
		} 
	\end{table}                                                   
	This paper considers one ground BS, which is equipped with an MEC server, which receives data from the devices through the UAV  to support the computation-oriented communications task. The BS is located at the center with the height of $H$ [m]. 
	Taking into consideration the mobility of the UAVs in the network to collect data and determine the optimal trajectory, we introduce the following mobility limitations \cite{nguyen2020uav}: 
	\begin{equation}
		\Vert  X_{m,t} - X_{m,(t-1)} \Vert ^2 \leq D^2, \hspace*{2 mm} \forall m, t,
		\label{eq1}
	\end{equation}
	\begin{equation}
		\Vert  X_{m,t} - X_{m',t} \Vert ^2 \geq D_\text{min} ,  \hspace*{2 mm}\forall m, m', t,
		\label{eq2}
	\end{equation}
	\begin{equation}
		\sqrt {(X_{m,t} - \hat{X}_{k}) ^2} \leq r_\text{max} , \hspace*{2 mm} \forall m, k, t,
		\label{eq3}
	\end{equation}
	where $ D=1 \cdot v_\text{max} $ for all the UAVs and $v_\text{max}$ is the maximum speed of the UAVs. $D_\text{min}$ is the minimum distance between two UAVs. Device $k$ is in coverage of UAV $m$, if the horizontal distance is less than or equal to $r_\text{max}$. Eq. (\ref{eq1}) in the above expressions shows the maximum distance that UAV $m$ can horizontally move in time slot $t$, (\ref{eq2}) prevents the UAVs collision, and (\ref{eq3})  guarantees that the devices stay in the coverage region of UAVs.
	%Fig. 2 illustrates the scheduling of offloading and processing task. \\
	%\begin{figure}[H]
	%	\centering
	%	\includegraphics[width=1\linewidth]{hh}
	%	\caption{Scheduling and process of computing each task.}
	%\end{figure} 
	\subsection{Computation and Communication Model} 
	There are two channels in our work: 1) channel between the devices and the UAVs, and 2) channel between the UAVs and the BS. The computation tasks are offloaded to the UAV-MEC server for some processing, and then the results are passed to the BS for further processing. Fig. \ref{fig 2} illustrates the multi-tasks procedure. Depending on the transmission time and CPU processing capacity, the waiting time for changing states is between 0 and a few time slots. Now we consider the following steps:
	\begin{figure}[ht!]
		\centering 
		\includegraphics[width=0.6\linewidth]{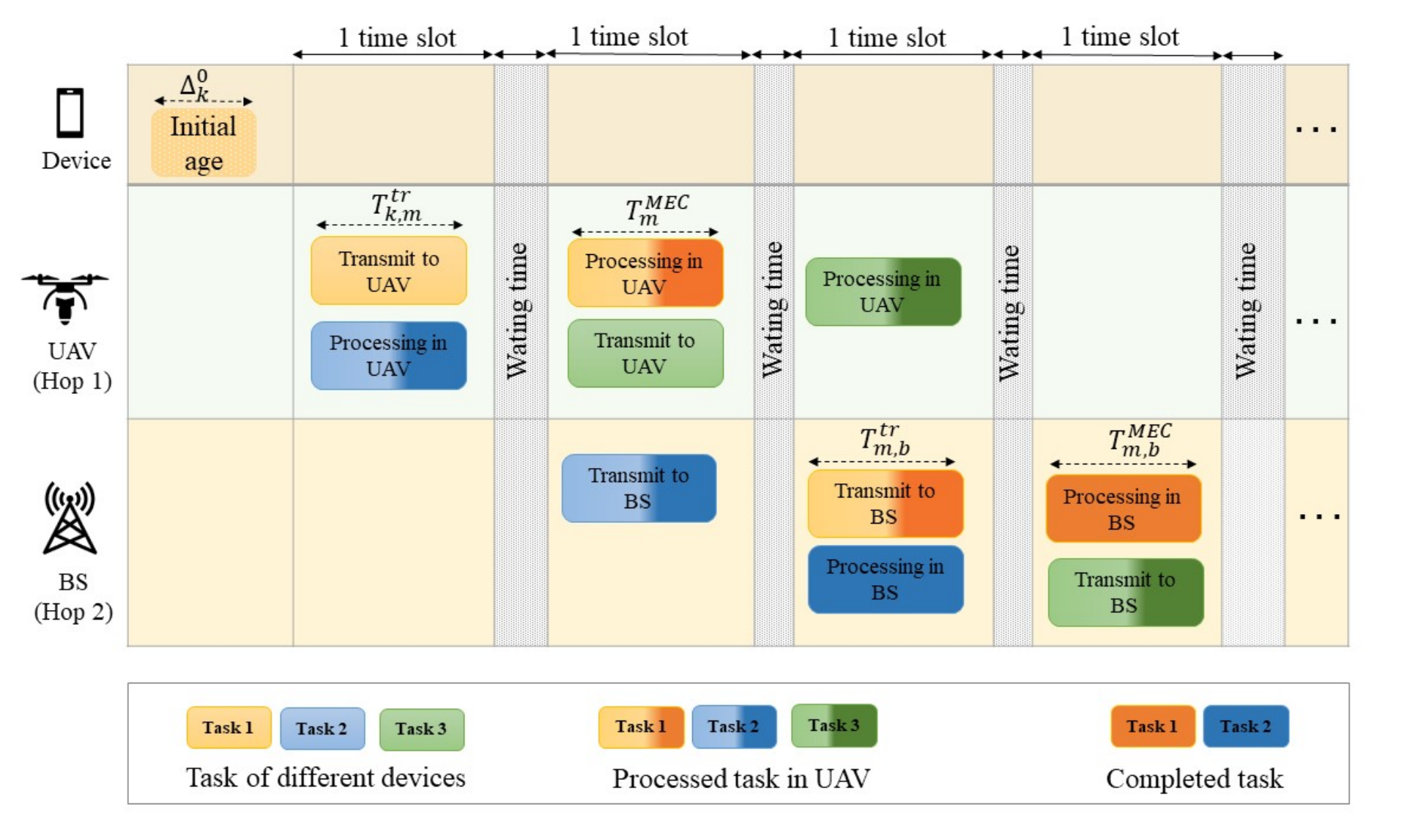}
		\caption{\color{black}{Transmission and procession of tasks: First, each task is received by the UAV (hop 1), and then part of the task is processed and sent to the BS (hop 2) at the appropriate time to continue processing.}}
		\label{fig 2}
	\end{figure} 
\\$\bullet $ \textbf{Step 1} The devices use orthogonal frequency division multiplexing (OFDM) channel with $F$ subcarriers. We define $\psi_{m,k,t}^{f}=1 $ to indicate that device $k$ decides to offload task to UAV $m$ at time slot $t$ on subcarrier $f$, otherwise $\psi_{m,k,t}^{f}=0 $. Each device can access the UAV via one subcarrier and can only connect to one UAV.  Then we have \cite{wang2021deep}
	\begin{equation}
		\sum_{m=0}^{M}\sum_{f=0}^{F} \psi_{m,k,t}^{f}\leq 1 ,\forall t, k\in \mathcal{K}.
	\end{equation}
	We assume  the communication channels from devices to the UAVs to be dominated by the LoS channel, and thus the channel gain between device $k$ and UAV $m$ at time slot $t$, is derived by\cite{yang2020multi}
	\begin{equation}
		g_{m,k,t}= \frac{\beta_{0}}{ h_m ^2+(x_{m,t}-\hat{x}_{k})^2+(y_{m,t}-\hat{y}_{k})^2},\hspace*{2 mm} \forall k, m, t,
	\end{equation}
	where $\beta_{0}$ represents the channel gain at the reference distance of 1 [m]. Moreover $ \rho_{k,t}$ denotes the transmit power of device $k$ at time slot $t$. The data rate in subcarriers $f$ can be expressed as 
	\begin{equation}
		R_{m,k,t}^{f}=B \log\left( 1+\frac{ \rho_{k,t} g_{m,k,t}}{\sigma^2}\right) , \hspace*{2 mm} \forall k, m,t,
	\end{equation}
	where $B$ denotes the channel bandwidth, $\sigma^2$ represents the white Gaussian noise variance. 
	The data offloading time to UAV $m$ at time slot $t$ is given by \cite{wang2020multi}
	\begin{equation}
		T_{k,m,t}^{f,tr}=\frac{\psi_{m,k,t}^{f} D_{k}}{R_{m,k,t}^{f}}  \hspace*{2 mm} \forall k, m,t.
	\end{equation}	
	In each UAV, the data offloading takes place over a maximum of one time slot, $ 	T_{k,m,t}^{f,tr}\leq 1$. Moreover, the execution time of the tasks at UAV $m$ in time slot $t$ can be expressed as \cite{ren2019collaborative}
	\begin{equation}
		T_{m,t}^\text{MEC}=\frac{\sum_{k}{\color{black}\psi_{m,k,t-1}^{f}} \Phi_{m,k,t} \lambda_{k} D_{k} F_{k}}{f_{m,t}^\text{MEC}} , \hspace*{2 mm} \forall  m,t,
	\end{equation}
	where $\lambda_{k} \in [0,1] $ is the task-breaking ratio, which indicates how much of a task is handled by the  UAV and $ f_{m,t}^\text{MEC}$ represents the computational capability of UAV $m$ at time slot $t$. The processing time in each UAV must be done in one time slot, i.e., $T_{m,t}^\text{MEC} \leq 1$. Moreover $ \Phi_{m,k,t}$ is a binary variable that indicates the task processing in UAV $m$ in time slot $t$ with $\Phi_{m,k,t}=1$, otherwise $ \Phi_{m,k,t}=0$.
	% To guarantee for each device $k$, transmission time to UAV and execution time in UAV not exceed duration time $\epsilon$, as well:
	%\begin{equation}
	%  T_{k,b}^{MEC}+T_{k}^{tr,b} \leq \epsilon, \forall k,
	%\end{equation}
	\\
	$\bullet $  \textbf{Step 2} Finally, the task is sent to the BS for further processing. Power domain non-orthogonal multiple access (PD-NOMA) is deployed with $L$ subcarriers for the communication between the UAVs and the BS. In the PD-NOMA-based network each UAV eliminates the other UAV’s signals by using the SIC approach, if $\lvert h_{m,b,t}^{l}\rvert ^ 2 \geq \lvert h_{m',b,t}^{l}\rvert ^ 2$, where $\lvert . \rvert  $ represents the absolute value, where the signal of UAV $m'$ is considered as noise.
	Therefore, the SINR of UAV $m$ on subcarrier $l$ at time slot $t$ is obtained by (\ref{sinr}), where $ h_{m,b,t}^{l} $ is the channel coefficient between the BS and UAV $m$ on subcarrier $l$, $p_{m,b,t}^{l}$ shows the transmit power at UAV $m$ to the BS on subcarrier $l$, $\zeta_{m,b,t}^{l} \in(0,1)$ is a binary variable that indicates the subcarrier allocation of UAV $m$ at time slot $t$, $(\sigma_{m}^{l})^2$ denotes the noise power of UAV $m$ on subcarrier $l$ \cite{moltafet2017comparison}.
	
		\begin{equation}
			\gamma_{m,b,t}^{l}= \frac{\zeta_{m,b,t}^{l}\arrowvert h_{m,b,t}^{l}\arrowvert ^2 p_{m,b,t}^{l}}{\sum_{\substack{m^{\prime} \neq m \\|h_{m',b,t}^{l}|^2 \leq |h_{m,b,t}^{l}|^2} } \arrowvert h_{m',b,t}^{l}\arrowvert ^2p_{m,b,t}^{l} \zeta_{m,b,t}^{l}+(\sigma_{m}^{l})^2}, \forall m, t, l,
			\label{sinr}
		\end{equation}

	%%\begin{figure*}[ht!]
	%	\begin{align}
	%		\gamma_{m,b,t}^{l}= \\ \nonumber &
	%		\frac{\zeta_{m,b,t}^{l}\arrowvert h_{m,b,t}^{l}\arrowvert ^2 p_{m,b,t}^{l}}{\arrowvert h_{m,b,t}^{l}\arrowvert ^2 \sum_{\substack{m^{\prime} \in \mathcal{M} \\ m^{\prime} \neq m}} p_{m^{\prime},b,t}^{l} \zeta_{m^{\prime},b,t}^{l}+(\sigma_{m}^{l})^2}, \forall m,
	%	\end{align}
	%%	\hrule
	%%\end{figure*}	
	Moreover, each subcarrier can be assigned to $L_l$ UAVs $\sum_{m\in\mathcal{M}} \zeta_{m,b,t}^{l} \leq L_l  $ and a total transmit power constraint for each UAV is
	\begin{equation}
		\sum_{l\in \mathcal{L}} \zeta_{m,b,t}^{l} p_{m,b,t}^{l} \leq P_{m}^\text{max} , \hspace*{2 mm} \forall m,t.
	\end{equation}
	The rate of UAV $m$ on subcarrier $l$  is formulated as $ \tilde{R}_{m,t}^{l}=\log (1+\gamma_{m,b,t}^{l} ),\forall m\in M$ and the sum data rate of UAV $m$ at time $t$ is denoted by
	\begin{equation}
		\tilde{R}_{m,t}=\sum_{l\in\mathcal{L}}\log (1+\gamma_{m,b,t}^{l}) , \hspace*{2 mm}\forall m, t.
	\end{equation}
	The total transmission time from UAV $m$ to the BS can be evaluated as
	\begin{equation}
		{\color{black} T_{m,b,t}^{l,tr}=\frac{\sum_{l}\sum_{k}\zeta_{m,b,t}^{l}z_{k,m,t}D'_{k}}{\tilde{R}_{m,t}} , \hspace*{2 mm}\forall m, t,}
	\end{equation}
	{\color{black} where $z_{k,m,t}$ is a binary variable, for which $z_{k,m,t}=1$, the task $k$ in UAV $m$ is selected for transmission to the BS in time slot $t$, otherwise $z_{k,m,t}=0$. $D'_k$ represents the input data size after processing in the UAV (in bits)}. The total transmission time from UAV $m$ to the BS is limited by one time slot, i.e., $T_{m,b,t}^{l,tr} \leq 1$. 
	The execution time of UAV $m$ in the BS can also be expressed as:
	\begin{equation}
		T_{m,b,t}^\text{MEC}=\frac{ \sum_{k} z_{k,m,t-1}j_{k,m,t}^{b}(1-\lambda_{k}) D_{k} F_{k}}{f_{m,b,t}^\text{MEC}}  , \hspace*{2 mm}\forall m, t,
	\end{equation}
	where $ f_{m,b,t}^\text{MEC} $ represents the computational capability of the BS, which can be allocated to UAV $m$ at time slot $t$, and $ j_{k,m,t}^{b}$ is a binary variable that shows the task processing in BS in time slot $t$. Moreover, the processing time in the BS must be less than one time slot, i.e., $T_{m,b,t}^\text{MEC} \leq 1$. 
	%To guarantee for each device $k$, transmission time to BS and execution time in BS not exceed duration time $\epsilon$, as well:
	%\begin{equation}
	%T_{k}^{tr,b}+T_{k,b}^{MEC} \leq \epsilon' , \forall k,
	%\end{equation}
	\subsection{Age of information}
	In order to determine the freshness of information, we consider AoI.  We formulate the AoI of the task in different hops. Each task has an initial age before being sent to the UAV, which is obtained by
	\begin{equation}
		\Delta_{k,t+1}^{0}= (1- \psi_{m,k,t}^{f})+ \Delta_{k,t}^{0},
	\end{equation}
	such that if the task is not sent to the UAV $(\psi_{m,k,t}^{f}=0)$, AoI will increase. 
	In this part, we formulate AoI of the task processed in two hops\cite{9637803}. For device $k$, let $\Delta_{k,t}^{m}$ and $\Delta_{k,t}^{b}$ be AoI of task $k$ at the UAV and at the BS at time slot $t$. $\Delta_{k,t}^{m}$ and $\Delta_{k,t}^{b}$ depend on how the scheduling is done.
	\\ When task $k$ arrives at the UAV for processing, $\Delta_{k,t}^{m}$ will be equal to AoI of the task in device plus one. In each time slot, AoI will increase by one until the task is inside the UAV and is not transmitted to the BS. Therefore, AoI of task $k$ at the UAV, $\Delta_{k,t+1}^{m}$, can be updated as \eqref{eq15},  where $t_{k}^0$ shows the generation time of task $k$.
		\begin{equation}
			\Delta_{k,t'+1}^{m}=
			\begin{cases}
				\Delta_{k,t'}^{0}+1	, & if\hspace*{2 mm}\hspace*{2 mm} \sum_{m=1}^{M}\sum_{f=1}^{F}\sum_{t=t_{k}^0}^{t'-1}\psi_{m,k,t}^{f}=0 \hspace*{2 mm}\hspace*{2 mm}\& \hspace*{2 mm}\hspace*{2 mm}\sum_{m=1}^{M}\sum_{f=1}^{F}\psi_{m,k,t'}^{f}=1   ,\\
				&\text{}
				\\
				\Delta_{k,t'}^{m}+1, &if \hspace*{2 mm}\hspace*{2 mm}\sum_{m=1}^{M}\sum_{f=1}^{F}\sum_{t=t_{k}^0}^{t'-1}\psi_{m,k,t}^{f}=1 \hspace*{2 mm}\hspace*{2 mm} \& \hspace*{2 mm}\hspace*{2 mm}\sum_{m=1}^{M}\sum_{l=1}^{L}\zeta_{m,b,t'}^{l} z_{k,m,t'}= 0.
			\end{cases}
			\label{eq15}
		\end{equation}	
	Similarly, the AoI in BS, $\Delta_{k,t+1}^{b}$, can be updated as \eqref{eq16}.
		\begin{equation}
			\Delta_{k,t'+1}^{b}=
			\begin{cases}
				\Delta_{k,t}^{m}+ 1 , & if\hspace*{2 mm} \sum_{m=1}^{M}\sum_{l=1}^{L} \sum_{t=t_{k}^0}^{t'-1}\zeta_{m,b,t}^{l} z_{k,m,t}= 0 \hspace*{2 mm}\hspace*{2 mm} \&\hspace*{2 mm} \hspace*{2 mm}\sum_{m=1}^{M}\sum_{l=1}^{L}\zeta_{m,b,t'}^{l}z_{k,m,t'}> 0,\\
				\\ 		\Delta_{k,t'}^{b}+ 1,  &if \hspace*{2 mm}\sum_{m=1}^{M}\sum_{l=1}^{L}\sum_{t=t_{k}^0}^{t'-1}\zeta_{m,b,t}^{l}z_{k,m,t} > 0 \hspace*{2 mm} \hspace*{2 mm}\& \hspace*{2 mm}\hspace*{2 mm} \sum_{m=1}^{M}j_{k,m,t'}^{b}=0\ .
			\end{cases}
			\label{eq16}
		\end{equation}
	The AoI is extended per time slot as long as the task is within the BS and is not yet fully processed.
	The average AoI depends on the data size, the allocated resources, required number of CPU cycles, data rate, and computing capacity of the MEC in the UAV and the BS for data processing.
	We minimize an objective function of a weighted sum average AoI via optimizing offloading location, resource allocation and the UAVs trajectories. The optimization problem in one time period $T$ can be expressed as follows:
	\begin{subequations}
		\begin{align}
			&\min_{\boldsymbol{\psi},\boldsymbol{\Phi},\boldsymbol{z}, \boldsymbol{\zeta},\boldsymbol {j} }\frac{1}{KT} \sum\limits_{t=0}^{T} \sum\limits_{k=1}^{K} \Delta_{k,t}^{m}\label{1:1}\\ 
			s.t :
			\label{C_1}&\sum_{m\in\mathcal{M}} \zeta_{m,b,t}^{l} \leq L_l,\\
			\label{C_2}&T_{k,m,t}^{f,tr}\leq 1,\forall k, t\\
			\label{C_3}&T_{m,t}^\text{MEC} \leq 1,\forall m, t\\
			\label{C_4}& T_{m,b,t}^{l,tr} \leq 1,\forall m, t\\
			\label{C_5}& T_{m,b,t}^\text{MEC} \leq 1,\forall m, t\\
			\label{C_6}&\sum\limits_{k\in\mathcal{K}}\Phi_{m,k,t} \lambda_{k} D_{k} F_{k}\leq f_{m,t}^\text{max} ,\forall m, k, t\\
			\label{C_7}&\sum\limits_{k\in\mathcal{K}} j_{k,m,t}^{b}(1-\lambda_{k}) D_{k} F_{k}\leq f_{m,b,t}^\text{max}, \forall m, t\\
			\label{C_8}& \psi_{m,k,t}^{f},\Phi_{m,k,t}  \in \{0,1\},\hspace*{2 mm} \forall k, m, t\\
			\label{C_9}& \zeta_{m,b,t}^{l},z_{k,m,t}, j_{k,m,t}^{b}  \in \{0,1\},\hspace*{2 mm} \forall k, m, t\\ \nonumber
			Eq. \label{C_10} & (1), (2), (3).
		\end{align}	
	\end{subequations}
	In the above formulated problem, constraint \eqref{C_1} denotes subcarrier constraint in PD-NOMA link, \eqref{C_2}, \eqref{C_3}, \eqref{C_4}, \eqref{C_5} indicate time limits for transmission and processing for each task and \eqref{C_6} shows the maximum CPU processes cycle for each UAV in time slot $t$, and \eqref{C_7} deals with the maximum CPU processes cycle, which the BS allocates to UAV $m$.
	\section{ MARL-based Resource Management Scheme}
	In this work, the computing resources and user association is coupled with each other. As a result of having both integer and continuous variables, the formulated problem is nonconvex, and the complexity of the problem is high. Since the action space is large, estimating state space is problematic, and solve the optimization problem globally is extremely hard. Furthermore, the problem grows in complexity as the number of devices covered by the UAVs increases. Using traditional methods to resolve the problem can be challenging, so it is better to use RL instead. In this work, the optimization problem is modeled with a MDP. We propose the RL method for solving the MDP problem and present a multi-agent solution.
	RL is based on actor-critic systems; the direct agent corresponds to states and actions, rather than providing a probability distribution across a discrete action space. As a result, MADDPG algorithms are good solutions. 
	Considering the collaboration between multiple agents in this article, we recommend using multi-agent methods. Q-learning and policy gradient as well as traditional reinforcement learning methods, are not suitable to multi-agent environments. Using two different types of agents, each of which having a unique set of states and actions, this weakness is further elaborated on in our approach. Therefore, our proposed method here is federated learning, which by sharing information between agents, the efficiency of the model is increased.
	\subsection{Problem Transformation}
	%There are several components to the proposed MADDPG algorithm. Denote $\mathcal{S}$ as the environment state space, for each state, the agents chooses association, resource allocation for processing, and power allocation action $a$ from the action space $\mathcal{A}$ according to the current policy $\pi$.
	%Then by doing $a$ in the environment,  will pay a reward $r$  to the agents for managing the policy updating until an optimal policy is obtained at time $t$.
	%\\Indeed, DDPG method employs two more techniques not present in the original DQN. first, it uses two target networks, because it adds balance to training, in brief, we are learning from predicted targets, and target networks are updated slowly, hence saving our estimated targets stable. second, it uses experience replay, by saving the components, and instead of learning only from new experiences, we learn from sampling all of our experience gained so far. we can express each part as
	We will describe the multi-agent environment here, including its associated states, actions, and rewards, and then we will discuss the multi-agent RL algorithm proposed here and relevant formulation.
	\\A multi-agent environment aims to maximize the policy functions of each agent, which can be described as follows \cite{parvini2021aoi}:
	\begin{equation}
		\max_{\pi_w} J_{w} (\pi _ {w}) , \hspace*{2 mm} w \in \left\lbrace \mathcal{M} ,b\right\rbrace \hspace*{2 mm}  ,\pi_w \in \Pi_\text{W},
	\end{equation}
	where $J_{w} (\pi _ {w})= \mathop{E} [\sum_{t=0}^{\infty} \gamma ^{t} \tilde{r}^{t+1}_{w} \vert s^{0}_{w}] $ is a conditional expectation, where $\mathop{E}\left[. \right]$ shows the statistical expectation. We denote  $\pi_{w} \in\left\lbrace \pi _{m}  \cup \pi_b \right\rbrace $ for agents's policy, where $\pi_m= \left\lbrace {\pi_1,...,\pi_\text{M}}\right\rbrace $ refers to the set of polies of the UAVs agent and $\pi_b $ refers to policy of the BS agent, and $\Pi_\text{W} $ is all possible policies for all agents. The goals of the agents policy are described in this section, as well as how each agent reacts to the UAV network  environment. Solving the optimization problem is the final goal. In this paper, due to the existence of two hops to calculate the AoI, the network in both hops possess separate actions, states, and rewards, and we have two type of agents.
	\\\textbf{State}: 
	The state of the environment $s \in \mathcal{S}$ at the time $t$ in UAV is expressed as
	\begin{equation}
		s_{t}^{U}= \left\lbrace \Delta_{k,t-1}^{m},g_{m,k,t},X_k,X_{m,t},V_{m}\right\rbrace, \hspace*{1 mm} U \in \mathcal{M}  .
	\end{equation}	
	The states of the UAV include all channel gains, all previous AoI, the velocity of the UAV, the user's location and the UAV's previous location.
	The state of the environment $s \in \mathcal{S}$ at the time $t$ in the BS is expressed as
	\begin{equation}
		s_{t}^{b}= \left\lbrace \Delta_{k,t-1}^{b}, h_{m,b,t}^{l}, I_{m,b,t}^{l}\right\rbrace  .
	\end{equation}
	The state of BS includes channel gain, all AoI in previous time slot and interference. The whole state is shown by:
	\begin{equation}
		S_{T}= \left\lbrace s_{t}^{U}, s_{t}^{b} \right\rbrace, \hspace*{1 mm} U \in \mathcal{M}.
	\end{equation}
	\textbf{Action}: We model the actions of UAVs as their associations with devices, CPU cycle allocations, velocities and transmit powers to the UAV. We relaxed the binary variable to be continuous in order to deploy the RL-based method. The binary variable is mapped to be between 0 and 1. The action selected by the UAV at time $t$ is given by
	\begin{equation}
		a_{t}^{U}=\left\lbrace {p_{m,b,t}^{l},\psi_{m,k,t}^{f},f_{m,t}^\text{MEC},V}\right\rbrace , \hspace*{1 mm} U \in \mathcal{M},
	\end{equation}
	and the action of BS at time slot $t$ is 
	\begin{equation}
		a_{t}^{b}=\left\lbrace {\zeta_{m,b,t}^{l},f_{m,b,t}^\text{MEC}}\right\rbrace,
	\end{equation}
	which includes the association with the UAV and CPU cycle allocation. The whole action is:
	\begin{equation}
		A_{T}=\left\lbrace {a_{t}^{U}, a_{t}^{b}}\right\rbrace, \hspace*{2 mm} U \in \mathcal{M} .
	\end{equation}
	\textbf{Reward}:
	The reward is a numerical value received by the agent from the environment and quantifies the degree to which the agent’s objective has been achieved. We must formulate a function correctly so that it can both represent the objective function and allow us to attain a faster and more stable convergence. We describe two types of reward functions in our work. First, we will define a reward for agents that collaborate to minimize the average age of tasks as part of an age-sensitive MEC system. The reward at time slot $t$ for each agent is expressed as
	\begin{equation}
		r_{t}^{U}= -k_{1}\Delta_{k,t+1}^{m}, \hspace*{1 mm}r_{t}^{U} \in J_{w},
	\end{equation}
	where $0 \leq k_{1}\leq 1$ is the reward coefficient. Second, we determine the global optimization of the network, where the following long-time reward is examined:
	\begin{equation}
		r_{t}^{b}= -k_{2}\Delta_{k,t+1}^{b}, \hspace*{1 mm}r_{t}^{b} \in J_{w},
	\end{equation}
	where $0 \leq k_{2}\leq 1$ is long-term reward coefficient.
	\subsection{Actor-Critic Network }
	Parameters of the actor and critic networks of the agents are $\theta_\pi= \left( W_\pi^{1},..., W_\pi^{L_\pi} \right) $ and $\phi_q= \left( W_q^{1},..., W_q^{L_q} \right) $ and the parameter set for the global critic is $\psi_g= \left( W_g^{1},..., W_g^{L_g} \right)$. The W's are the neural network's weight matrices, and the number of nodes in the hidden layers determines the dimensions of each matrix. Accordingly, $L_\pi$, $L_q$, and $L_g$ are the number of hidden layers in the actor and critic systems of agents, and the global critic. The global critic’s Q-function $Q_{ \psi}^{g}$ are parameterized by ${ \psi} $, and the Q-functions $ Q_{\phi_m}^{m}$ are parameterized by ${\phi_m}$.
	% Policy agents $\pi_{w}$, the Q-function Q, and global critic's function  
	The MADDPG for each agent is expressed as
	\begin{align}
		&\nabla_{\theta_w} J_{w}= \mathop{E} [\nabla_{\theta_w} \pi^{w} \left(a_w \vert  s_w \right) 
		\nabla_{A_w} Q_{w}^{\pi}(s,a) \vert_{A_w=\pi_w(S_w)} ] ,
	\end{align}
	where $\theta_w$ is parameterized for agent's policies $\pi_w$ and the Q-function $Q_{w}^{\pi}(s,a)$ is the global action value function that estimates Q-value based on the actions and states of the agents.
	Here we present the policy gradient for each agent, which consists of two critic networks:
	\begin{align}
		&\nabla_{\theta_w} J_{w}= \mathop{E_{s,a\sim D}}\left[  \nabla_{\theta_b} \pi^{b} \left(a_b \vert  s_b \right)  \nabla_{a_{b}} Q_{ \psi}^{g}(s,a) \right]
		+\mathop{E_{s_m,a_m\sim D}}\left[  \nabla_{\theta_m} \pi^{m} \left(a_m \vert  s_m \right)  \nabla_{a_m} Q_{\phi_m}^{m}(s,a) \right]  ,
	\end{align}
	where the first term depicts the global critic, which takes action and state from BS and calculates the global reward. The second term depicts the local critic, which takes action and state from UAVs.  We update the loss function as follows:
	\begin{equation}
		{L(\psi)}= \mathop{E_{s,a,r,s'}} \left[ (Q_{ \psi}^{g}(s,a)-y_g)^2 \right] ,
	\end{equation} 
	where $y_g$ is the target value and can be written as
	\begin{equation}
		y_g=r_g+ \gamma Q_{ \psi}^{g'}(s',a') \vert_{a_b'=\pi_b'(s_b')}  .
	\end{equation} 
	We also update the other loss function as
	\begin{equation}
		{L(\phi_m)^m}= \mathop{E_{s_m,a_m,r_m,s'_m}} \left[ (Q_{ \phi_m}^{m}(s_m,a_m)-y_l^m)^2 \right], 
	\end{equation}
	where $y_l^m$ is described as
	\begin{equation}
		y_l^m= r_l^m+ \gamma Q_{ \phi_m}^{m'}(s'_m,a'_m) \vert_{a_m'=\pi_m'(s_m')}.
	\end{equation}
	Eventhough the present approach can produce decent results, there is still the problem of overestimation and suboptimal strategies in Q-functions caused by the function estimate errors. Therefore, we replace the twin delayed deterministic policy gradient with the global critic \cite{parvini2021aoi}.  %% maqale parvini reference
	\begin{align}
		&\nabla_{\theta_w} J_{w}= \mathop{E_{s,a\sim D}}\left[  \nabla_{\theta_b} \pi^{b} \left(a_b \vert  s_b \right)  \nabla_{a_{b}} Q_{ \psi_1}^{g_1}(s,a) \right]+
		\mathop{E_{s_m,a_m\sim D}}\left[  \nabla_{\theta_m} \pi^{m} \left(a_m \vert  s_m \right)  \nabla_{a_m} Q_{\phi_m}^{m}(s,a) \right]  .
	\end{align}
	Accordingly, the twin global critics are updated as follows:
	\begin{equation}
		{L(\psi_j)}= \mathop{E_{s,a,r,s'}} \left[ (Q_{ \psi_j}^{g_j}(s,a)-y_g)^2 \right] ,
	\end{equation} 
	where $y_g$ is the target value and can be written as
	\begin{equation}
		y_g=r_g+ \gamma \min Q_{ \psi_j'}^{g_j}(s',a') \vert_{a_w'=\pi_w'(s_w')}.
	\end{equation}  
Algorithm 1 presents our proposed RL-based method.
	\begin{algorithm}[h]
		\tiny
		\DontPrintSemicolon
		\renewcommand{\arraystretch}{0.5}
		\caption{FRL for AoI scheduling}
		\label{algorithm}
		Initiate environment, generate UAVs, BS and devices \;  
		$\mathbf{Inputs}$: Enter number of $A_T$, $S_T$\;
		%	$\mathbf{Outputs}$: The current action $a_t$.\;
		Initialize all, global critic networks $Q^{f}{\phi_1}$ and $Q^{f}{\phi_1}$, target global critic networks $Q^{\prime f}{\phi_1}$ and $Q^{\prime f}{\phi_1}$ and agents policy and critic networks.\;
		\For{t=1 to T}
		{
			\For{n = 1 : $N$ }
			
			\hspace*{2 mm}{\If{n$\geq$2}{Aggregate the neural networks weights.\;
					Update the weights.
					\hspace*{2 mm}	\hspace*{2 mm}	}\hspace*{2 mm}\For{each agnet $w$ }
				{\eIf{w=0}{Observe state $s^{b}_t$ and take action $a^{b}_t$}{Observe state $s^{u}_t$ and take action $a^{u}_t$}
					%	{ Observe state $s^{f}_t$ and take action $a^{f}_t$ }
					$\mathbf{S}{T}=\left[s^{b}_t,s^{b}_t\right], \quad \mathbf{A}{T}=\left[a^{b}_t,a^{u}_t\right]$.\;
					Receive global and local rewards, $\tilde{r}{G,t}$ and $\tilde{\mathbf{r}}{t}^{f}$\; Store $\left(\mathrm{s}{t}, \mathbf{a}{t}, \tilde{\mathbf{r}}{t}^{f}, \tilde{r}{G,t}, \mathbf{s}_{t+1}\right)$ in replay buffer $\mathcal{D}$
				}
				\hspace*{2 mm}	Sample minibatch of size $\mathrm{S},\left(\mathrm{s}^{j}, \mathbf{a}^{j}, \mathbf{r}{g}^{j}, \mathbf{r}{\ell}^{j}, \mathbf{s}^{\prime} j\right)$, from
				replay buffer $\mathcal{D}$\; \hspace*{2 mm}Set $y_{g}^{j}=r_{g}^{j}+\gamma \min {i} Q{\psi_{i}^{\prime}}^{g_{i}}\left(\mathbf{s}^{\prime}, \mathbf{a}^{\prime} j\right)$\;
				\hspace*{2 mm}	Update global critics by minimizing the loss:\;
				\begin{equation*}
					\mathcal{L}\left(\psi_{i}\right)=\frac{1}{S} \sum_{j}\left\{\left(Q_{\psi_{i}}^{g_{i}}\left(\mathrm{~s}^{j}, \mathbf{a}^{j}\right)-y_{g}^{j}\right)^{2}\right\}.
				\end{equation*}
				\;
				\hspace*{2 mm}Update target parameters: $\psi_{i}^{\prime} \leftarrow \tau \psi_{i}+(1-\tau) \psi_{i}^{\prime}$\;
				\hspace*{2 mm}	\If{episode mod $d$}{Train actor and critic nerwork\;
					\For{for each agent f }{  episode mod $d$ 
						$\mid \begin{aligned}&\mathcal{L}\left(\phi_{i}\right)=\frac{1}{S} \sum_{j}\left\{\left(Q_{\phi_{i}}^{i}\left(s_{i}^{j}, a_{i}^{j}\right)-y_{i}^{j}\right)^{2}\right\}. \\&\text { Update local actors: } \\&\nabla J_{\theta_{i}} \approx \frac{1}{S} \sum_{j}\left\{\nabla_{\theta_{i}} \pi_{i}\left(a_{i} \mid s_{i}^{j}\right) \nabla_{a_{i}} Q_{\psi_{1}}^{g_{1}}\left(\mathbf{s}^{j}, \mathbf{a}^{j}\right).\right.\end{aligned}$
						$\left.\nabla_{\theta_{\imath}} \pi_{i}\left(a_{i} \mid s_{i}^{j}\right) \nabla_{a_{i}} Q_{\phi_{i}}^{i}\left(s_{i}^{j}, a_{i}^{j}\right)\right\}$\;
						Update target networks parameters:
						$\left[\begin{array}{l} \theta_{i}^{\prime} \leftarrow \tau \theta_{i}+(1-\tau) \theta_{i}^{\prime}, \\ \phi_{i}^{\prime} \leftarrow \tau \phi_{i}+(1-\tau) \phi_{i}^{\prime}.\end{array}\right.$
		}}}}
	\end{algorithm}	
	\subsection{Federated Model}
	%Multi-agent learning scenarios require the sharing of knowledge between different agents to achieve the global optimum. Additionally, transmission and processing of the observation will consume too many communication and computation resources. Thus, as to overcome these difficulties, we propose a federated mechanism for the above framework, where agents' actor parameters are shared across all agents at every ${E_f} $ learning epoch, and federated updating is carried out throughout all agents. Therefore, each  agent maintains the factors with weight $w$ and combines the other factors, as follows \cite{zhu2021federated}:
The proposed RL-based solution is multi-agent-based and relies on a central weight-aggregator to manage and boost cooperation among agents via information sharing between the agents as well as having a local critic that estimates the local expected reward. In other words, the better we can foster cooperation between several agents, the more promising RL can learn the environment and lead us to stable convergence to achieve satisfactory results. As opposed to local information, such as state and action, an FRL agent transmits the weights of its actor-networks to a central server at every ${E_f} $ learning epoch. The server collects these weights and runs them through a pre-set algorithm, then aggregates them and sends them back to the agents. The central server's aggregation rules are acquired by\cite{zhu2021federated}
	\begin{equation}
		\boldsymbol{\theta}^{t+1}=\boldsymbol{\theta}^{t}\cdot\mathbf{\Omega}.
	\end{equation}
	At the $t$-th learning epoch, we have $ \boldsymbol{\theta}^{t}=\left\lbrace \boldsymbol{\theta}^{t}_1,..., \boldsymbol{\theta}^{t}_{M} \right\rbrace $, where $M$ denotes all the agents' parameters at the t-th learning epoch, and  vector $\mathbf{\Omega}$ is calculated as follows:	
	\begin{equation*}
		\boldsymbol{\Omega} = 
		\begin{pmatrix}
			w & {1-w}\over{M-1} & \cdots & {1-w}\over{M-1}  \\
			{1-w}\over{M-1} & w & \cdots & {1-w}\over{M-1}  \\
			\vdots  & \vdots  & \ddots & \vdots  \\
			{1-w}\over{M-1} & {1-w}\over{M-1} & \cdots &w 
		\end{pmatrix}.
	\end{equation*}	
Agents now assess actions that reduce average AoI in the whole network to increase policy function, which is directly related to the other agents actions and behavior. By adhering to the multi-agent case, the agents do not necessarily need to take the other agents states and actions, and this is the superiority we can gain through multi-agent algorithms. However, the Achilles heel in MADDPG is that as the agents act somehow independently, the environment becomes highly susceptible to the agents actions. Introducing the federated model in the proposed algorithm can slightly mitigate this problem, which is inevitable in the conventional MADDPG. The proposed framework increases the cooperation among the agents, avoids the learning struggles and leads to a better performance.
	%The federated model allows each agent to communicate with each other. The system improves its communication efficiency by transmitting only the parameters of lightweight actor nets rather than sharing input states. Further, convergence  in the federated model  is faster. 
	\subsection{Computational Complexity}
	Investigating the computational complexity is critical to the utility of the proposed algorithms. In essence, the complexity depend on four parameters, i) the number of trainable parameters, ii) the total number of neural networks used in the algorithms, iii) the computational complexity, iv) the communication overhead between the UAVs and the central server. We can sufficiently understand their relevance and practicality through this encyclopedic standpoint.
	In MADDPG method, all observations and actions of the agents are considered as inputs. Several parameters need to be defined, including action for the UAVs and the BS (denoted by $a$ and $a'$), and state for the UAVs and the BS (denoted by $s$ and $s'$). As a result, the number of trainable parameters for the MADDPG algorithm is $\mathcal{O}\left( n\left( a+s\right) +\left( a'+s'\right)\right) $, where $n$ is the number of UAV agents and one BS as an agent, respectively. Both MADDPG and FRL algorithms follow the same pattern, the FRL method does not increase the number of trainable parameters. The number of networks in conventional MADDPG is $2 ×\left( n \left( \underline{1}_Q + \underline{1}_A\right) + \left( \underline{1}_{Q'} + \underline{1}_{A'}\right) \right) $. The reason for multiplying by two is the extra network as the target. Critic and actor networks for UAVs are represented by $\underline{1}_{Q}$ and $\underline{1}_{A}$, and critic and actor networks for the BS are represented by $\underline{1}_{Q'}$ and $\underline{1}_{A'}$ respectively. The number of nodes in the neural network in the FRL algorithm is similar to MADDPG.
	\subsection{Overhead}
	A performance metric often overlooked is the communication overhead, especially when communications are established using a MARL framework. MARL frameworks rely heavily on agents' communication.
	This information exchange is essential for stabilizing the learning process and encouraging cooperation between agents. Still, it is necessary to keep the overhead as low as possible. 
	As a means of calculating and analyzing this information, we assume that each matrix element can be decoded as a 16-bit binary vector. So, we use a type of float16 in Numpy. The overall communication overhead is provided in the following for each suggested framework. 
	The overhead in MADDPG is equal to $ \left( 16 \times (M-1) \times s^{u}+ 16 \times 1 \times M \times S \right) $ KB, and the overhead in FRL is equal to $\left( 16 \times L_{\pi}+16 \times M \times L_{\pi} \right)$ KB.
	Consequently, the proposed method has less overhead than the MADDPG method. Table \ref{overhead} compares the overhead of FRL and MADDPG method.
	\begin{table}[ht!]
		\centering
		\caption{Comparison of the overhead between MADDPG and FRL}
		\label{overhead}
		\scalebox{0.7}{
			\begin{tabular}{|c|c|}
				\hline
				Algorithm & Overhead                               \\ \hline
				MADDPG   & $\left( 16 \times (M-1) \times s^{u}+ 16 \times 1 \times M \times S \right)$  bits\\ \hline
				FRL         & $\left( 16 \times L_{\pi}+16 \times M \times L_{\pi} \right)$  bits                                   \\  \hline
			\end{tabular}		} 
	\end{table}
	\section{Simulation}
	This section shows the average AoI at the UAVs and the BS, using the FRL method, and compares the results with the MADDPG method. There are 5 UAVs, 15 devices on the ground, and the environment is limited to an area of 200 $\times$ 200 [m] during the simulation. A slot and a frame duration are 1 ms and 100 ms, respectively. Device generates a task with communication demand $D_k \in \left[ 20,50\right] $ KB and computation requirement is $F_K \in  \left[ 3 \times 10^{9},3 \times 10^{10}\right] $ processes cycle. In the following neural network, hyper-parameter is summarized in Table \ref{parameter}.
	At the beginning of each slot, the velocity of each UAV is determined, and up to the next slot, the UAV hovers based on it. Then during each slot within each frame, the UAV assignment and power allocation for users are handled. Furthermore, the RL selects the best user to connect with in order to minimize AoI. It is important to mention that during the initial phase of learning, the RL does not yield satisfactory results, and the value of reward function is small. As the system begins to learn and adapt to the environment, i.e., better estimation of the system dynamics is rendered, the reward function increases. In other words, the best power levels are selected for each UAV so that the interference between the UAVs becomes as small as possible and UAVs can deliver better quality of service. Finally, a proper trajectory design for UAVs is decided by RL so that they act more flexibly and precisely leading the tasks to be delivered at their earliest convenience.
	%In Fig. \ref{fig 3}, the rewards achieved per episode in the training steps using the FRL and MADDPG methods in different subchannels are presented. In the first 200 episodes of the FRL and MADDPG method, since the parameters of all agents are globally initialized by TensorFlow based on the state in the first episode, the rewards are small and change rapidly over time. While the training continues, the actors and critics change their parameters for measuring and targeting the networks to gradually achieve optimal policies and also the state-action functions corresponding to optimal policies. The rewards are fairly stable after episode 200.
	\begin{figure}[ht!]
		\centering
		\includegraphics[width=0.5\linewidth]{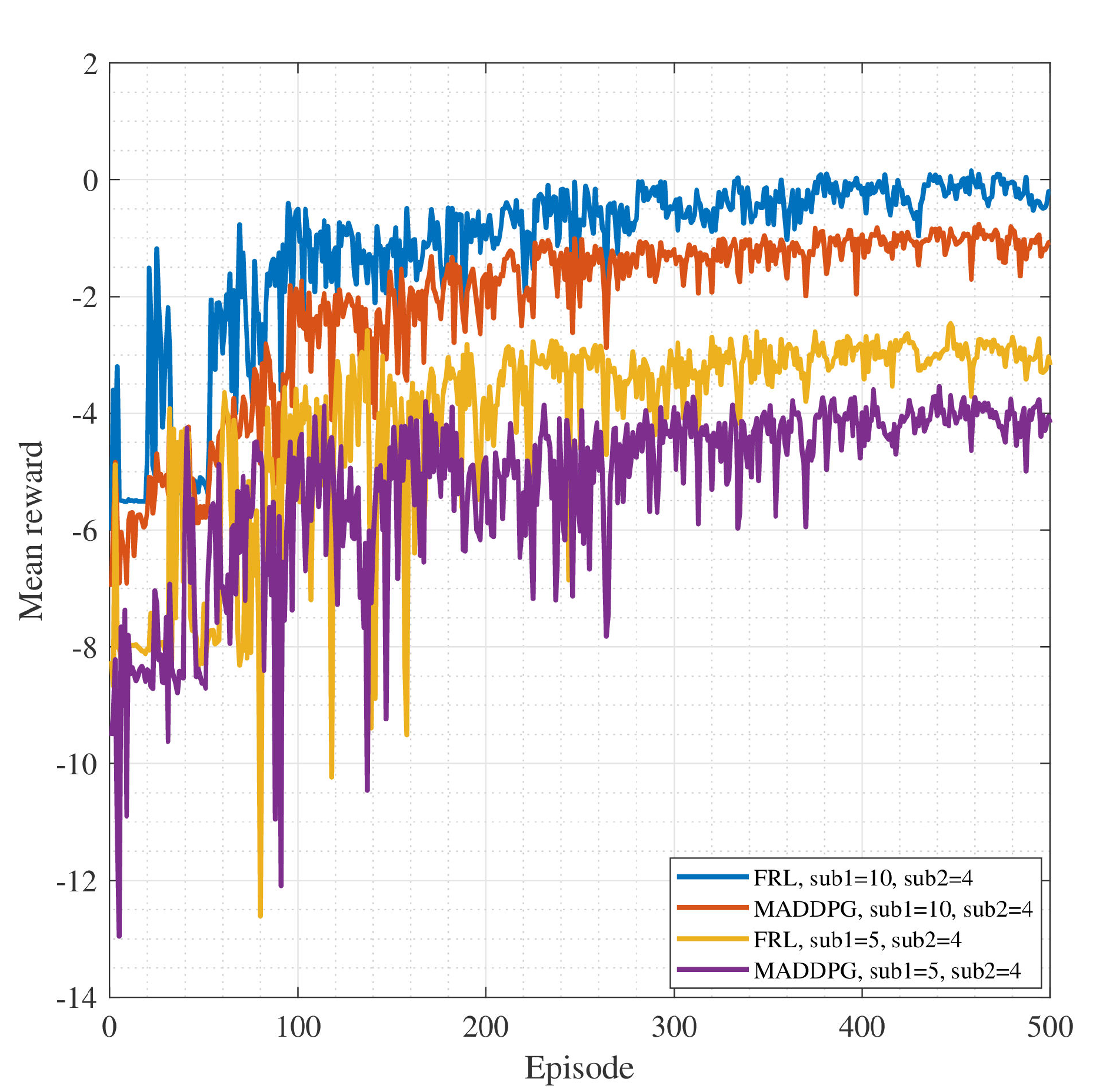}
		\caption{Rewards achieved per episode during training.}
		\label{fig 3}
	\end{figure} 
	In the following, we present the results obtained from the numerical simulation. First, we describe the effect of the number of subchannels and various multiple access methods, then examine the reward. Finally, we analyze the performance of the system model without UAV.
	\begin{table}[ht!]
		\centering
		\caption{Network Parameters}
		\label{parameter}
		\scalebox{0.7}{
			\begin{tabular}{|ll|ll|}
				\hline
				Neural networks hyper-parameters & Value & Neural networks hyper-parameters& Value                                     \\ \hline
				Number of local actor networks hidden layers    &3/2048/1024/512 & Critic/Actor networks learning rate                 &0.001/0.0001  \\
				Number of local critic networks hidden layers         & 2/1024/512  & Discount factor         & 0.99          \\ 
				Number  of global critic hidden layers            & 2/2048/1024  & Number of episodes                   & 500   \\ 
				Experience replay buffer size              & 500000      & 	Number of iterations per episode      & 100   \\ 
				Mini batch size                 & 64  & Target networks soft update parameter        & 0.0005         \\  
				\hline
			\end{tabular}
		} 
	\end{table}
	\subsubsection{Trajectory}
\begin{figure}[ht!]
		\centering
		\subfigure[The UAVs trajectories for 9 device.]{\includegraphics[width=6cm,height=6cm]{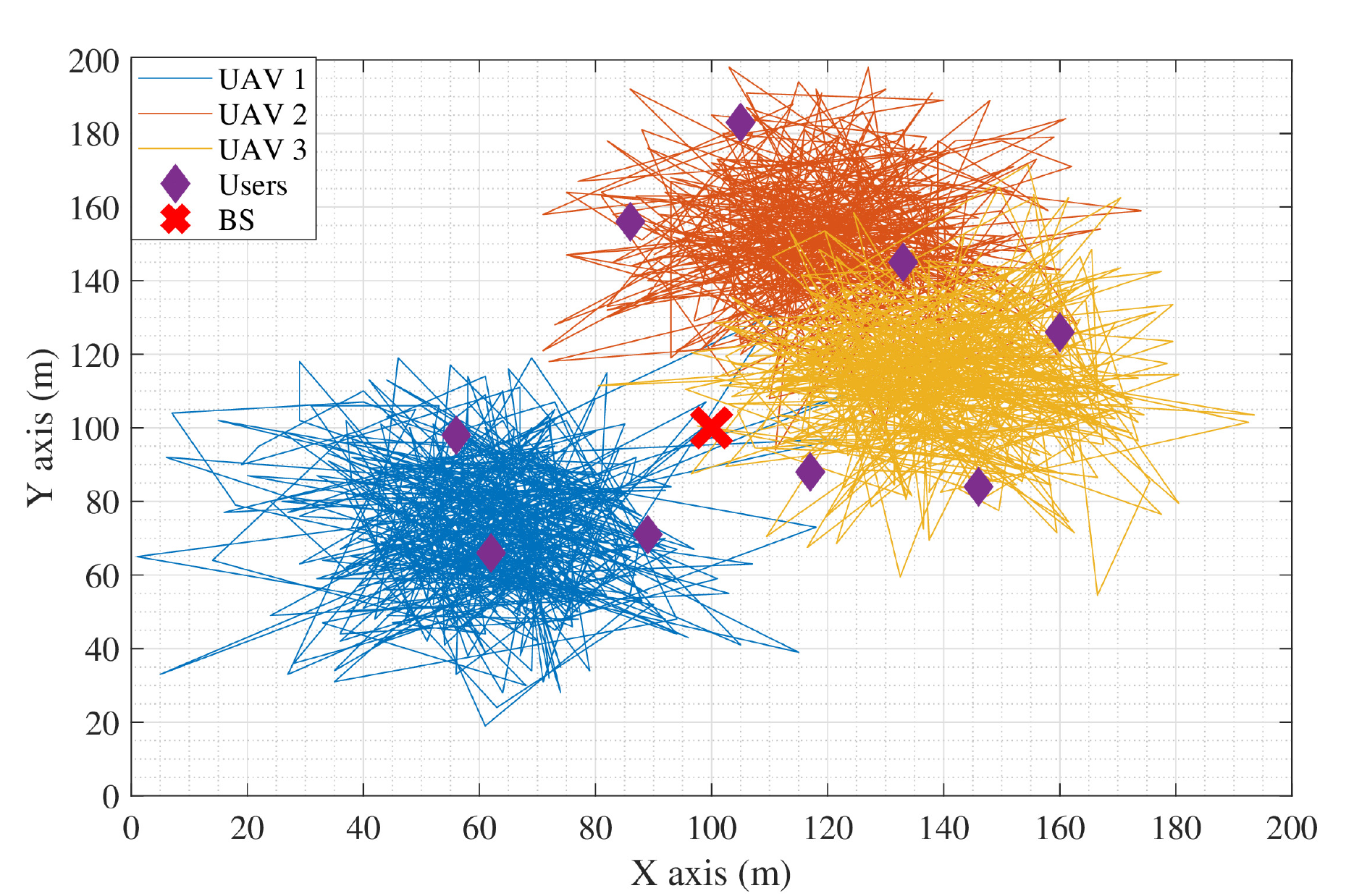}} \,\,
		\subfigure[The UAVs trajectories for 18 device.]{\includegraphics[width=6cm,height=6cm]{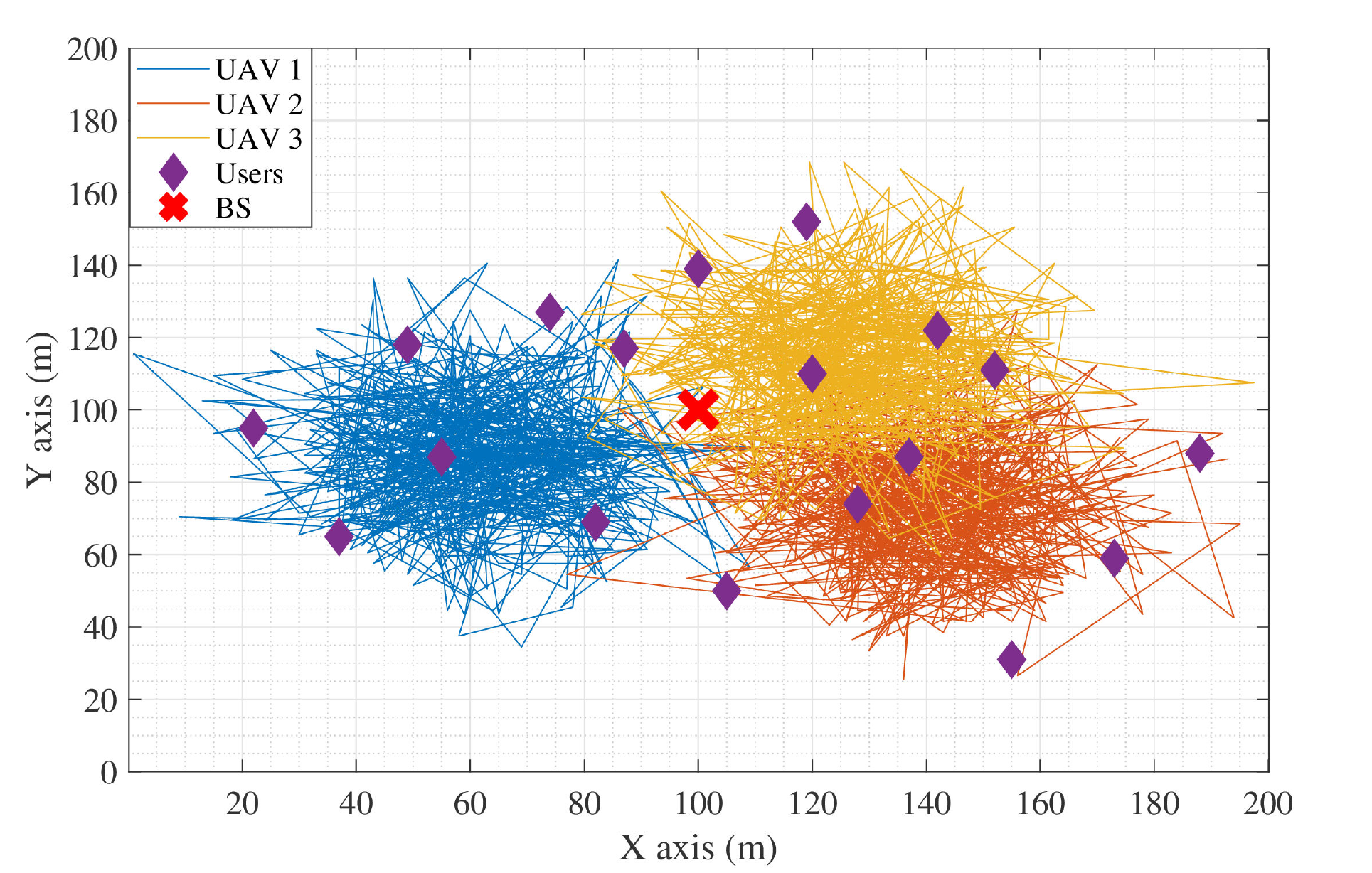}}  \,\,	
		\caption{The trajectories of UAVs in the proposed scheme.}\label{Main_Fig}
		\label{traj}
	\end{figure}

In Fig. \ref{traj}, we illustrate the UAVs trajectories in the coverage area and the given locations of the users and the located BS. As the number of devices increases, the mobility of the UAVs to collect data increases to keep the information freshness at each receiver, which results in more overlap in the areas covered.
	\subsubsection{Baseline Model}
In order to evaluate the proposed system model, we prepare scenarios as system model baselines for comparison. In our system model, we assumed two scenarios: \textbf{Network without} UAV, which meansinstead of having partially processing, the whole task is transmitted directly to BS. \textbf{Through a pure OFDMA}, we considered OFDMA between the UAVs and the BS, unlike the main approach, which is using NOMA.\\	
	In the solution part, we assume \textbf{MADDPG} as the baseline algorithm.
Similar to the single-agent actor-critic model, each agent has its own network of actors and critics. Actor networks take into account the current state of the agent and recommend an action based on it. Nevertheless, the critic network is quite different from the usual single-agent DDPG. In this algorithm, all the UAVs and the BS behave as an agent.
	\subsubsection{Effect of the number subchannels and users on AoI}
\begin{figure}
		\centering
		\subfigure[Mean AoI in UAVs with different devices]{\includegraphics[width=.4\textwidth]{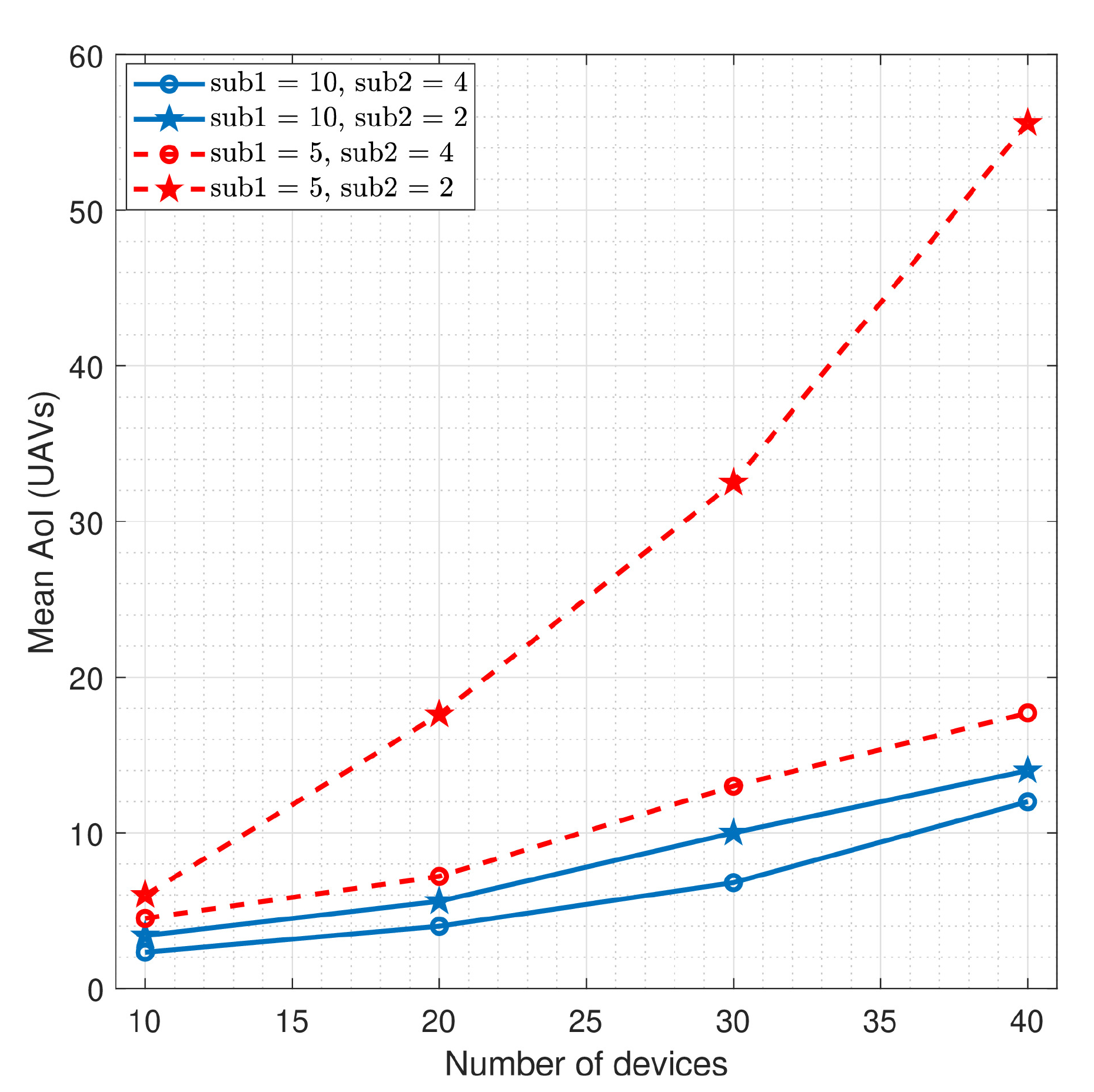}} \,\,
		\subfigure[Mean AoI in BS with different devices]{\includegraphics[width=.4\textwidth]{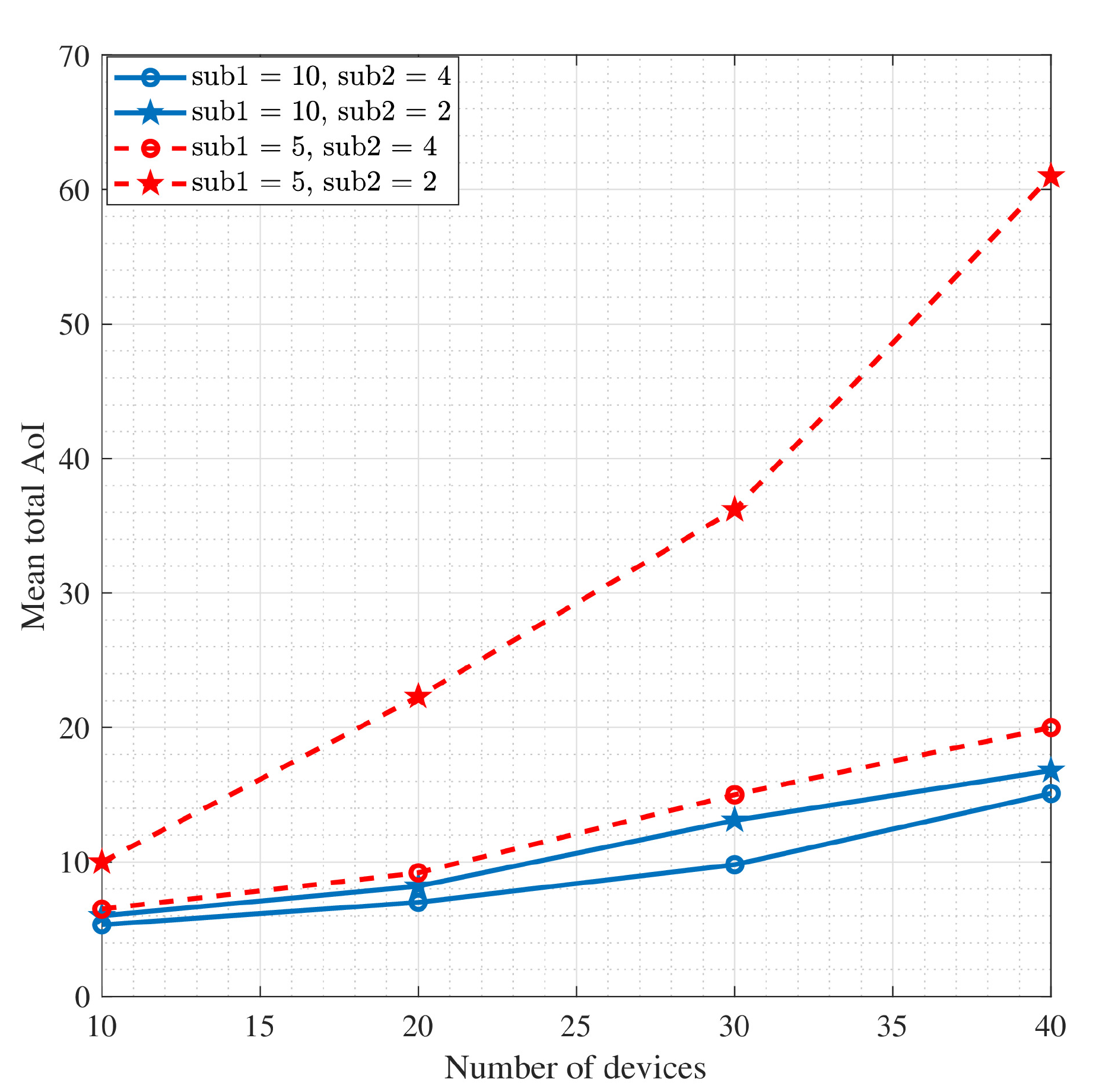}}\,\,
		\caption{Effect of the number of subchannels and devices on AoI.}\label{Main_Fig}
		\label{fig 5}
	\end{figure}

	Due to the network's limited capacity, we see that the information freshness status deteriorates  as the number of users increases. Given current resources, there can only be a finite number of users scheduled to transmit their status data during the given time slots. Therefore, waiting times for submitting new updates are generally longer, which increases the AoI.
	Moreover, as the number of subcarrier increases, we observe a dramatic decrease in AoI both at UAV and BS in Fig. \ref{fig 5}.
	\subsubsection{Network without UAV}
\begin{figure}
		\centering
		\subfigure[Network without UAV and with UAV.][\label{UAV}]{\includegraphics[width=.4\textwidth]{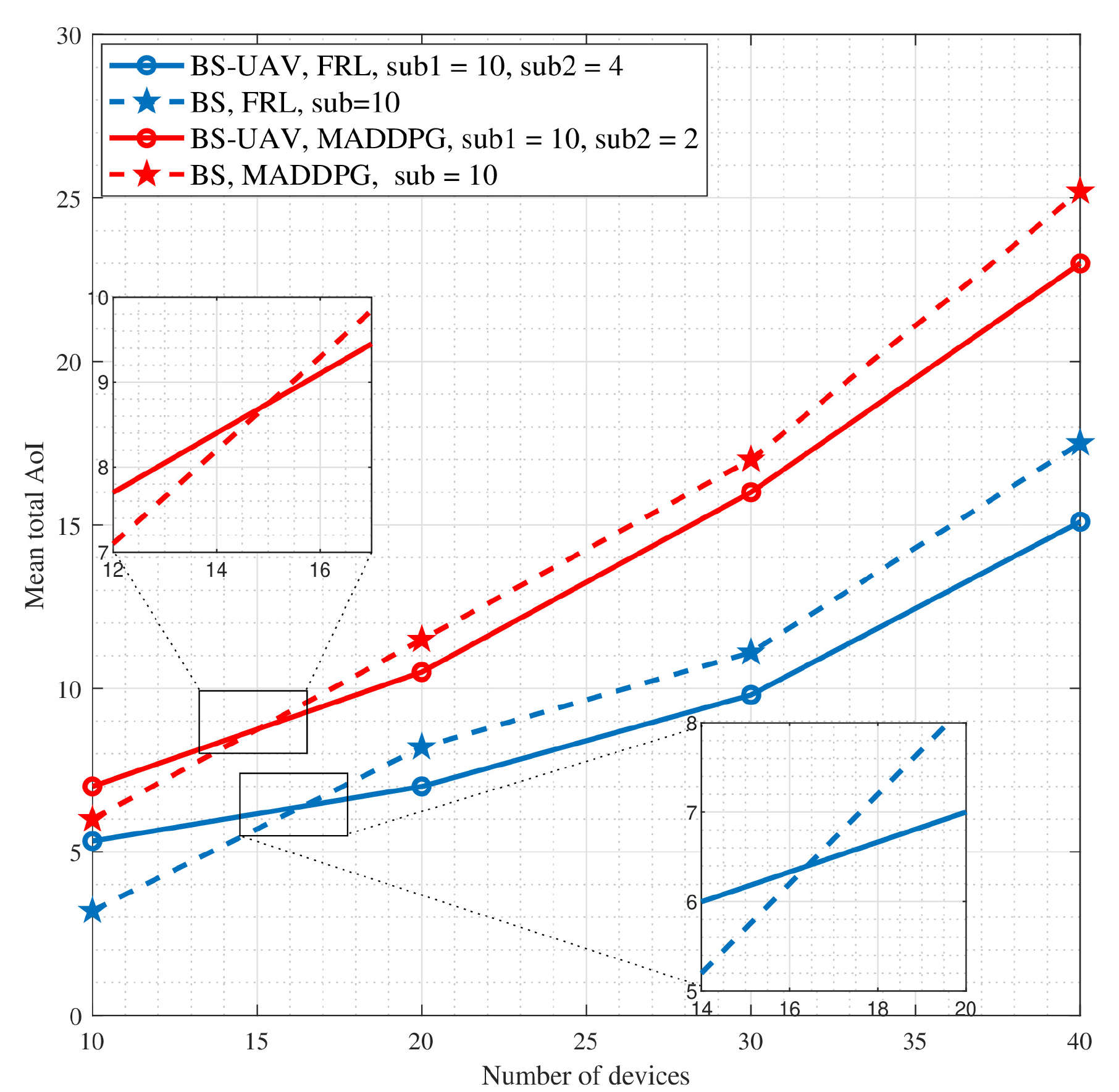}}\,\,
		\subfigure[Mean AoI in fixed subchannel with different multiple access.][\label{subchannel}]{\includegraphics[width=.4\textwidth]{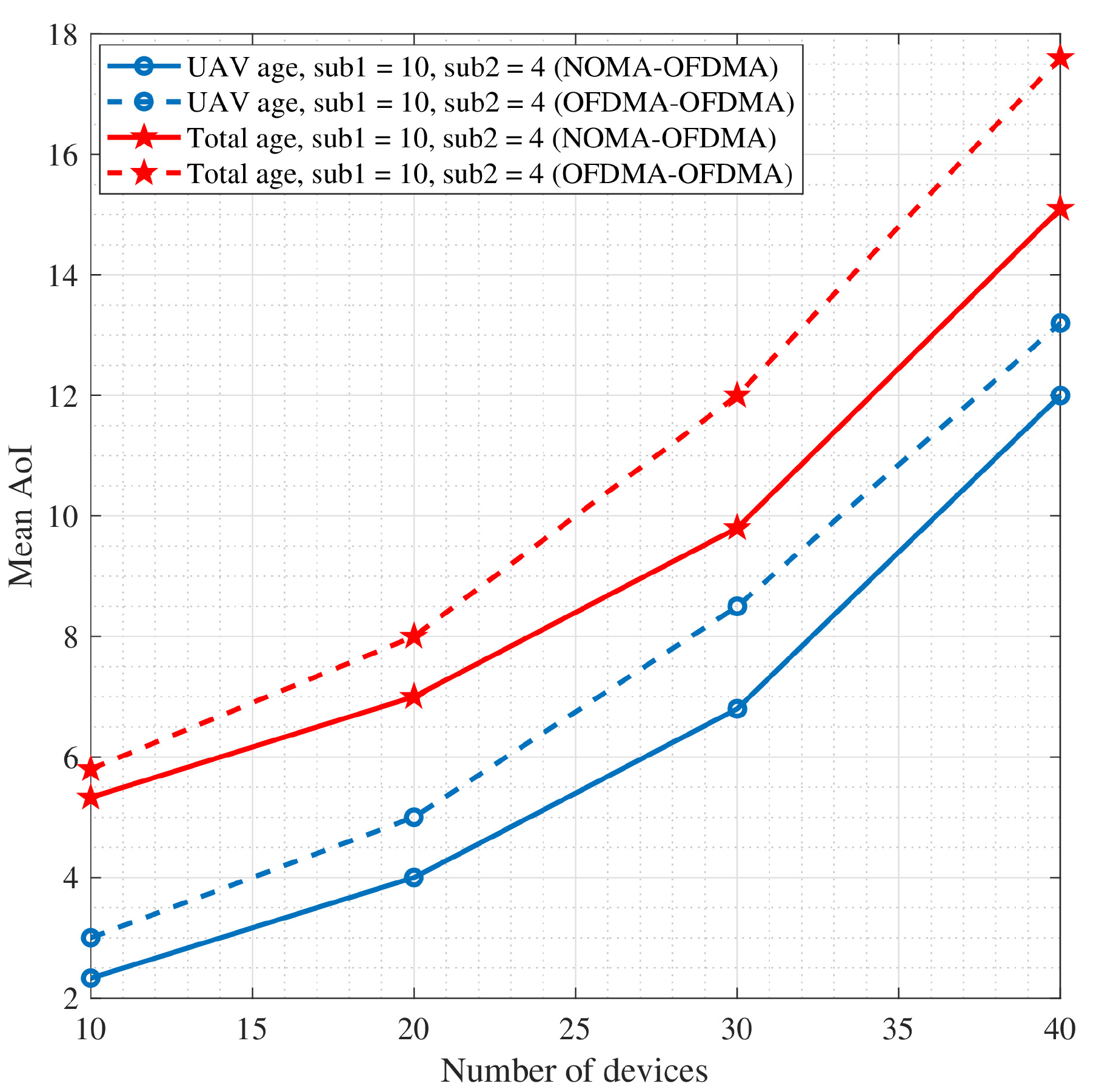}}\,\,
		\caption{Effect of deployment of UAVs and different multiple access on AoI.}
	\end{figure}

	Efficacy of collaboration between UAV and BS is yielded from the simulation result, especially  as the number of devices increases. As seen in Fig. \ref{UAV},  in both FRL and MADDPG algorithms, the network without UAV perform better when the number of devices is smaller than about 16. As the number of devices exceed 16, UAVs and BS collaboration improves the AoI. In-line with to the increase in the number of tasks to be processed, network processing capacity needs to increase too. Lack of processing capacity increases the waiting time for processing, which affects the freshness of information. Therefore, the simultaneous use of MEC in UAVs and BS reduces network processing load and thus reduces AoI.
	\begin{figure*}[ht!]
		\centering
		\subfigure[Mean AoI in BS with FRL and MADDPG algorithm under (OFDMA-NOMA) multiple access]{\includegraphics[width=.4\textwidth]{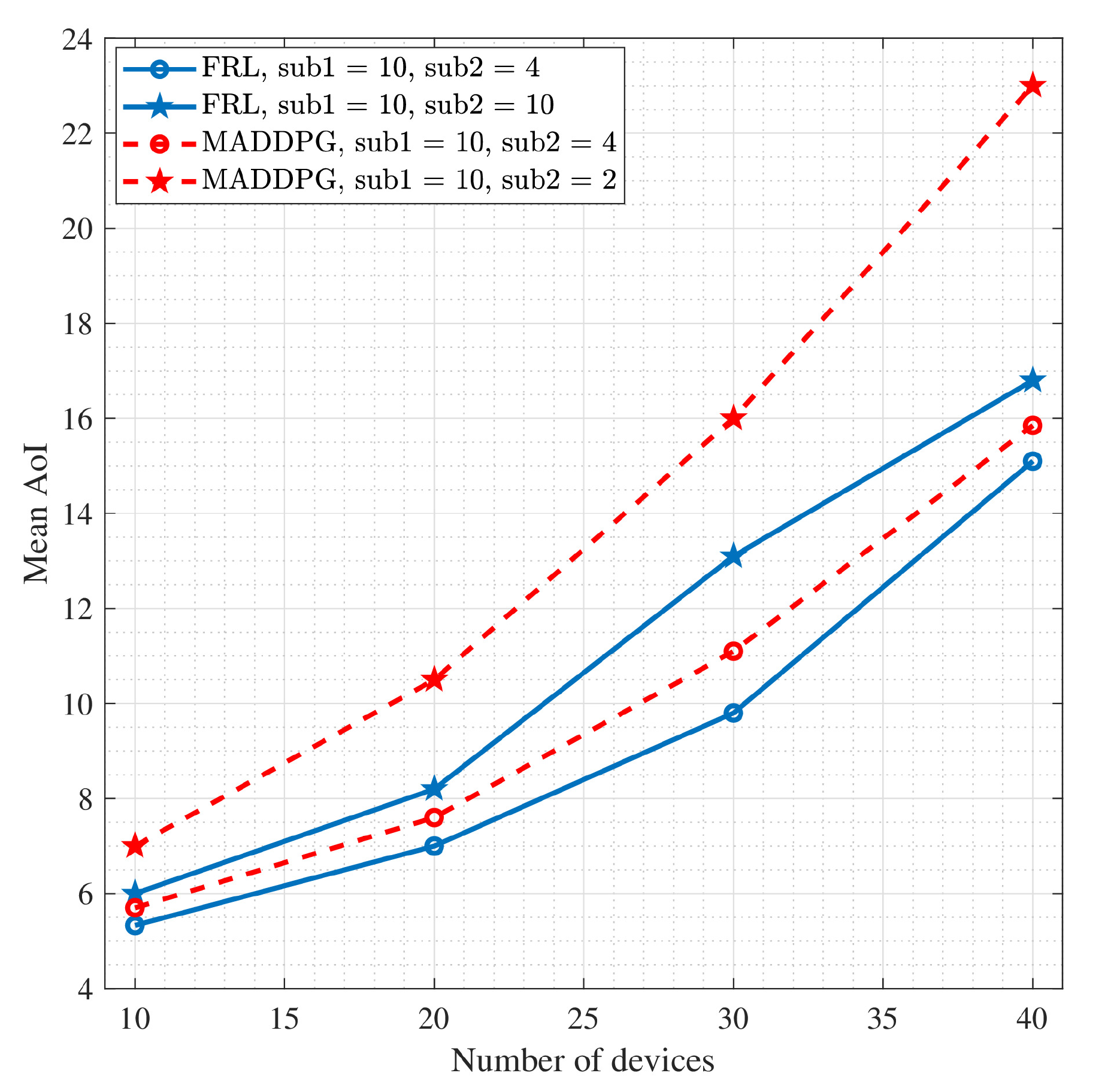}}\,
		\subfigure[Mean AoI in BS with FRL and MADDPG algorithm under (OFDMA-OFDMA) multiple access]{\includegraphics[width=.4\textwidth]{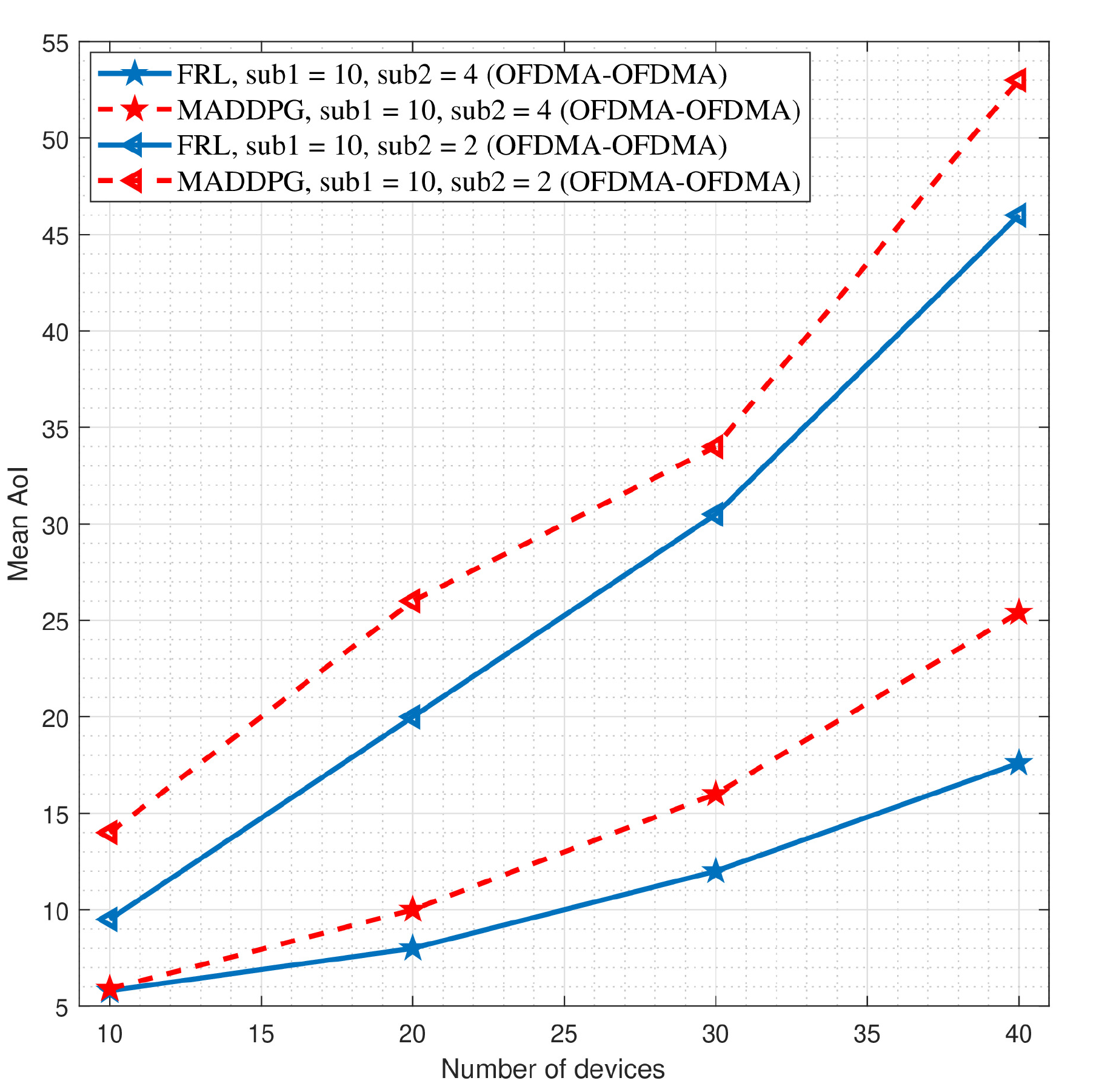}}\,\,
		\caption{Comparison of AoI between FRL vs MADDPG. }\label{Main_Fig}
		\label{fig 7}
	\end{figure*}
	\subsubsection{Comparison between different multiple access scheme}
	Despite the complexity of NOMA, the use of NOMA-OFDMA multiple access performs better than OFDMA-OFDMA and improves AoI in the network, as shown in Fig. \ref{subchannel}. In the NOMA method, two UAVs can use the same subcarrier to send data, in other words, network capacity increases. Consequently, network performance is enhanced, leading to improved AoI.
	\subsubsection{Comparison between FRL and MADDPG}
	As mentioned, increasing the number of subcarriers improves the AoI. In Fig. \ref{fig 7}, the increase in number of subcarriers between the UAVs and the BS, along with different multiple access schemes using the two algorithms, is investigated. In either case, the FRL method performs better than MADDPG. The FRL has a weights update center, which improves agent efficiency, and thus it outperforms MADDPG. 
	\section{Conclusion}
In this paper, an FRL-based multi-task transmission and resource allocation method was developed for UAV-MEC systems, aiming at minimizing the two-hop AoI of UAV-BS while guaranteeing the task delivery to BS. Under the proposed MARL algorithm, with the central weight-aggregator, a group of UAVs can simultaneously learn how to maximize the cooperative global reward and their individual local rewards while sharing their collective model with other UAVs. Furthermore, UAVs are designed using a proper trajectory by RL so that they can act in a more precise and flexible manner in order to collect tasks. Via such a mechanism, we reveal that the proposed two-hop scheme is exceptionally vigorous and effective in boosting UAVs to improve system-level implementation, although the UAVs and BS independently select their assignment,  power levels, and trajectory. In conclusion, we validated the performance and effectiveness of the proposed FRL method through encyclopedic simulation. As a future work, an in-depth extension of the proposed system model can be conducted for mobile-users scenarios for the UAV system. A further promising direction will be examining the spectrum sharing scenarios in UAV networks.
	\bibliographystyle{ieeetr}
	\bibliography{R}	

\begin{thebibliography}{10}

\bibitem{9285214}
S.~F. Abedin, M.~S. Munir, N.~H. Tran, Z.~Han, and C.~S. Hong, ``Data
  {F}reshness and {E}nergy-{E}fficient {UAV} {N}avigation {O}ptimization: {A}
  {D}eep {R}einforcement {L}earning {A}pproach,'' {\em IEEE Transactions on
  Intelligent Transportation Systems}, vol.~22, no.~9, pp.~5994--6006, 2021.

\bibitem{abedin2018resource}
S.~Abedin, M.~G.~R. {A}lam, S.~A. {K}azmi, N.~H. {T}ran, D.~{N}iyato, and h.~S.
  {H}ong, ``Resource allocation for ultra-reliable and enhanced mobile
  broadband {IoT} applications in fog network,'' {\em \scriptsize IEEE
  Transactions on Communications}, vol.~67, no.~1, pp.~489--502, 2019.

\bibitem{li2018uav}
B.~Li, Z.~{F}ei, and Y.~{Z}hang, ``{UAV} communications for {5G} and {B}eyond:
  {R}ecent {A}dvances and {F}uture {T}rends,'' {\em \scriptsize IEEE IoTs
  Journal}, vol.~6, no.~2, pp.~2241--2263, 2019.

\bibitem{mozaffari2019tutorial}
M.~Mozaffari, W.~Saad, M.~Bennis, Y.-H. Nam, and M.~Debbah, ``A tutorial on
  {UAVs} for wireless networks: {A}pplications, challenges, and open
  problems,'' {\em IEEE communications surveys \& tutorials}, vol.~21, no.~3,
  pp.~2334--2360, 2019.

\bibitem{yang2020multi}
L.~{Y}ang, H.~{Y}ao, J.~{W}ang, C.~{J}iang, A.~{B}enslimane, and Y.~{L}iu,
  ``Multi-{UAV}-enabled load-balance mobile-edge computing for {IoT}
  networks,'' {\em \scriptsize IEEE IoTs Journal}, vol.~7, no.~8,
  pp.~6898--6908, 2020.

\bibitem{mao2017survey}
Y.~{Mao}, C.~{You}, J.~{Zhang}, K.~{Huang}, and K.~B. {Letaief}, ``A survey on
  mobile edge computing: {The} communication perspective,'' {\em \scriptsize
  IEEE Communications Surveys Tutorials}, vol.~19, no.~4, pp.~2322--2358, 2017.

\bibitem{chen2020minimizing}
M.~Chen, Y.~{X}iao, Q.~{L}i, and K.-c. {C}hen, ``Minimizing age-of-information
  for fog computing-supported vehicular networks with deep {Q-learning},'' in
  {\em \scriptsize IEEE International Conference on Communications}, (Dublin,
  Ireland), pp.~1--6, July 2020.

\bibitem{yates2021age}
R.~D. {Yates}, Y.~{Sun}, D.~R. {Brown}, S.~K. {Kaul}, E.~{Modiano}, and
  S.~{Ulukus}, ``Age of information: {An} introduction and survey,'' {\em
  \scriptsize IEEE Journal on Selected Areas in Communications}, vol.~39,
  no.~5, pp.~1183--1210, 2021.

\bibitem{chen2020multiuser}
H.~Chen, Q.~{Wang}, Z.~{Dong}, and N.~{Zhang}, ``Multiuser scheduling for
  minimizing age of information in uplink {MIMO} systems,'' in {\em \scriptsize
  IEEE/CIC International Conference on Communications in China}, (Chongqing,
  China), pp.~1162--1167, November 2020.

\bibitem{abdel2018ultra}
M.~K. Abdel-{Aziz}, C.~{Liu}, S.~{Samarakoon}, M.~{Bennis}, and W.~{Saad},
  ``{U}ltra-reliable low-latency vehicular networks: {T}aming the age of
  information tail,'' in {\em \scriptsize IEEE Global Communications
  Conference}, (Arab Emirates), pp.~1--7, December 2018.

\bibitem{ei2021multi}
N.~Ei, S.~W. {Kang}, M.~{Alsenwi}, Y.~K. {Tun}, and C.~S. {Hong},
  ``{M}ulti-{UAV}-{A}ssisted {MEC} {S}ystem: {J}oint {A}ssociation and
  {R}esource {M}anagement {F}ramework,'' in {\em \scriptsize International
  Conference on Information Networking}, (Jeju Island), pp.~213--218, Jan 2021.

\bibitem{alsenwi2020uav}
M.~{Alsenwi}, Y.~K. {Tun}, S.~{Raj}~{Pandey}, N.~N. {Ei}, and C.~{Seon}~{Hong},
  ``{UAV}-assisted multi-access edge computing system: {A}n energy-efficient
  resource management framework,'' in {\em \scriptsize International Conference
  on Information Networking}, (Barcelona), pp.~214--219, Jan 2020.

\bibitem{yang2019energy}
Z.~{Yang}, C.~{Pen}, K.~{Wang}, and M.~{Shikh}-{Bahaei}, ``Energy efficient
  resource allocation in {UAV}-enabled mobile edge computing networks,'' {\em
  \scriptsize IEEE Transactions on Wireless Communications}, vol.~18, no.~9,
  pp.~4576--4589, 2019.

\bibitem{peng2020ddpg}
H.~{P}eng and X.~{Shen}, ``{DDPG}-based resource management for
  {MEC/UAV}-assisted vehicular networks,'' pp.~1--6, 2020.

\bibitem{wang2020multi}
L.~{W}ang, C.~{W}ang, { K}ezhi~and{ P}an, W.~{X}u, N.~{A}slam, and L.~{H}anzo,
  ``{M}ulti-{A}gent {D}eep {R}einforcement {L}earning {B}ased {T}rajectory
  {P}lanning for {M}ulti-{UAV} {A}ssisted {M}obile {E}dge {C}omputing,'' {\em
  \scriptsize IEEE Transactions on Cognitive Communications and Networking},
  vol.~7, no.~1, pp.~73--84, 2021.

\bibitem{hu2020cooperative}
J.~{Hu}, H.~{Zhang}, L.~{Song}, R.~{Schober}, and H.~{Poor}, ``Cooperative
  internet of {UAVs}: {D}istributed trajectory design by multi-agent deep
  reinforcement learning,'' {\em \scriptsize IEEE Transactions on
  Communications}, vol.~68, no.~11, pp.~6807--6821, 2020.

\bibitem{li2019minimizing}
W.~Li, L.~{W}ang, and A.~{F}ei, ``{M}inimizing packet expiration loss with path
  planning in {UAV}-assisted data sensing,'' {\em \scriptsize IEEE Wireless
  Communications Letters}, vol.~8, no.~6, pp.~1520--1523, 2019.

\bibitem{tong2019uav}
P.~{T}ong, J.~{L}iu, X.~{W}ang, B.~{B}ai, and H.~{D}ai, ``{UAV}-enabled
  age-optimal data collection in wireless sensor networks,'' in {\em
  \scriptsize IEEE International Conference on Communications Workshops},
  pp.~1--6, 2019.

\bibitem{8570843}
M.~A. {Abd-Elmagid} and H.~S. {D}hillon, ``Average {P}eak
  {A}ge-of-{I}nformation {M}inimization in {UAV}-{A}ssisted {IoT} {N}etworks,''
  {\em IEEE Transactions on Vehicular Technology}, vol.~68, no.~2,
  pp.~2003--2008, 2019.

\bibitem{yi2020deep}
M.~{Y}i, X.~{W}ang, J.~{L}iu, Y.~{Z}hang, and B.~{B}ai, ``Deep {R}einforcement
  {L}earning for {F}resh {D}ata {C}ollection in {UAV}-assisted {IoT}
  networks,'' pp.~716--721, July 2020.

\bibitem{liu2018age}
J.~{L}iu, X.~{W}ang, B.~{B}ai, and H.~{D}ai, ``Age-optimal trajectory planning
  for {UAV}-assisted data collection,'' in {\em \scriptsize IEEE Conference on
  Computer Communications Workshops}, pp.~553--558, 2018.

\bibitem{hu2020aoi}
H.~{Hu}, K.~{X}iong, G.~{Qu}, Q.~{N}i, P.~{Fan}, and K.~B. {L}etaief,
  ``{AoI}-minimal trajectory planning and data collection in {UAV}-assisted
  wireless powered {IoT} networks,'' {\em \scriptsize IEEE Internet of Things
  Journal}, vol.~8, no.~2, pp.~1211--1223, 2021.

\bibitem{samir2020age}
M.~{Samir}, C.~{Assi}, S.~{Sharafeddine}, D.~{Ebrahimi}, and A.~{Ghrayeb},
  ``Age of information aware trajectory planning of {UAVs} in intelligent
  transportation systems: {A} deep learning approach,'' {\em \scriptsize IEEE
  Transactions on Vehicular Technology}, vol.~69, no.~11, pp.~12382--12395,
  2020.

\bibitem{wu2021uav}
F.~{Wu}, H.~{Zh}ang, J.~{Wu}, Z.~{H}an, and L.~{Poor}, { H}
  {V}incent~and{S}ong, ``{UAV}-to-device underlay communications: {A}ge of
  information minimization by multi-agent deep reinforcement learning,'' {\em
  \scriptsize IEEE Transactions on Communications}, vol.~69, no.~7,
  pp.~4461--4475, 2021.

\bibitem{zhang2020age}
S.~{Z}hang, H.~{Z}hang, Z.~{H}an, H.~V. {Poor}, and L.~{ S}ong, ``Age of
  information in a cellular {I}nternet of {UAVs}: Sensing and communication
  trade-off design,'' {\em \scriptsize IEEE Transactions on Wireless
  Communications}, vol.~19, no.~10, pp.~6578--6592, 2020.

\bibitem{zhou2019deep}
C.~{Zhou}, H.~{He}, P.~{Yang}, F.~{Lyu}, W.~{Wu}, N.~{Cheng}, and X.~{Shen},
  ``Deep {RL}-based trajectory planning for {AoI} minimization in
  {UAV}-assisted {IoT},'' in {\em \scriptsize 11th International Conference on
  Wireless Communications and Signal Processing}, (Xi'an, China), pp.~1--6, Oct
  2019.

\bibitem{zhu2021federated}
Z.~Zhu, S.~Wan, P.~Fan, and K.~B. Letaief, ``Federated multiagent actor--critic
  learning for age sensitive mobile-edge computing,'' {\em IEEE Internet of
  Things Journal}, vol.~9, no.~2, pp.~1053--1067, 2021.

\bibitem{han2021age}
J.~{Han}, {Rui and Wang}, J.~{Bai}, {Lin and Liu}, and J.~{Choi}, ``{Age} of
  information and performance analysis for {UAV}-aided {IoT} systems,'' {\em
  \scriptsize IEEE IoTs Journal}, vol.~8, no.~19, pp.~14447--14457, 2021.

\bibitem{choudhury2021aoi}
R.~{Han}, J.~{Wang}, L.~{Bai}, J.~{Liu}, and J.~{Choi}, ``{AoI}-minimizing
  {S}cheduling in {UAV}-relayed {IoT} {N}etworks,'' {\em \scriptsize IEEE IoTs
  Journal}, vol.~8, no.~19, pp.~14447--14457, 2021.

\bibitem{abd2019deep}
M.~A. Abd-Elmagid, A.~Ferdowsi, H.~S. Dhillon, and W.~Saad, ``Deep
  {Reinforcement} {Learning} for {Minimizing} {Age}-of-{Information} in
  {UAV}-{Assisted} networks,'' in {\em \scriptsize 2019 IEEE Global
  Communications Conference}, (Waikoloa, HI, USA), pp.~1--6, Dec 2019.

\bibitem{kadota2018optimizing}
I.~{K}adota, A.~{S}inha, and E.~{M}odiano, ``{O}ptimizing age of information in
  wireless networks with throughput constraints,'' in {\em \scriptsize IEEE
  Conference on Computer Communications}, (Honolulu, HI, USA), pp.~1844--1852,
  April 2018.

\bibitem{wang2021deep}
L.~Wang, K.~Wang, C.~Pan, W.~Xu, N.~Aslam, and A.~Nallanathan, ``Deep
  reinforcement learning based dynamic trajectory control for {UAV}-assisted
  mobile edge computing ({Early Access}),'' {\em \scriptsize IEEE Transactions
  on Mobile Computing}, pp.~1--1, 2021.

\bibitem{nguyen2020uav}
M.~D. {N}guyen, T.~M. {H}o, L.~B. {L}e, and A.~{G}irard, ``{UAV} trajectory and
  sub-channel assignment for {UAV} based wireless networks,'' in {\em
  \scriptsize IEEE Wireless Communications and Networking Conference, 2020},
  (Seoul, Korea (South)), pp.~1--6, May 2020.

\bibitem{ren2019collaborative}
J.~{Ren}, G.~{Yu}, Y.~{He}, and G.~Y. {Li}, ``{C}ollaborative cloud and edge
  computing for latency minimization,'' {\em \scriptsize IEEE Transactions on
  Vehicular Technology}, vol.~68, no.~5, pp.~5031--5044, 2019.

\bibitem{moltafet2017comparison}
M.~{Moltafet}, N.~M. {Yamchi}, M.~R. {Javan}, and P.~{Azmi}, ``Comparison study
  between {PD-NOMA} and {SCMA},'' {\em \scriptsize IEEE Transactions on
  Vehicular Technology}, vol.~67, no.~2, pp.~1830--1834, 2018.

\bibitem{9637803}
B.~Choudhury, V.~K. {S}hah, A.~{F}erdowsi, J.~H. {R}eed, and Y.~T. {H}ou,
  ``{AoI}-minimizing {Scheduling} in {UAV}-relayed {IoT} {Networks},'' in {\em
  \scriptsize IEEE 18th International Conference on Mobile Ad Hoc and Smart
  Systems}, (Denver, CO, USA), pp.~117--126, Oct 2021.

\bibitem{parvini2021aoi}
M.~Parvini, M.~R. Javan, N.~Mokari, B.~A. Arand, and E.~A. Jorswieck, ``{AoI}
  {Aware} {Radio} {Resource} {Management} of {Autonomous} {Platoons} via
  {Multi} {Agent} {Reinforcement} {Learning},'' in {\em \scriptsize 17th
  International Symposium on Wireless Communication Systems}, (Berlin,
  Germany), pp.~1--6, Sept 2021.

\end{thebibliography}
\end{document}